\documentclass[sort&compress,10pt]{iopart}
\pdfoutput=1
\usepackage{iopams}
\usepackage{xspace}
\usepackage{bm}
\usepackage{ifthen}
\usepackage{algorithm}
\usepackage{graphicx}
\usepackage[usenames, dvipsnames]{color}
\usepackage{subcaption}
\usepackage{siunitx}
\usepackage{paralist}
\usepackage{cite}
\usepackage[scaled=0.9]{helvet}      
\usepackage{mathptmx}    
\usepackage[colorlinks=true,linkcolor=blue,citecolor=blue,urlcolor=blue,pdfborder={0 0 0}]{hyperref}

\newcommand{\potfit}{\emph{potfit}\xspace}
\newcommand{\kimpotfit}{\emph{KIM-potfit}\xspace}

\begin{document}

\title[A KIM-compliant potfit for fitting sloppy interatomic potentials]{A {KIM}-compliant \potfit for fitting sloppy interatomic potentials: {A}pplication to the {EDIP} model for silicon}

\author{Mingjian~Wen$^{1}$, Junhao~Li$^{2}$, Peter~Brommer$^{3}$, Ryan~S.~Elliott$^{1}$, James~P.~Sethna$^{2}$ and Ellad~B.~Tadmor$^{1}$\footnote{Corresponding author.}}
\ead{tadmor@umn.edu}

\address{$^{1}$ Department of Aerospace Engineering and Mechanics, University of Minnesota, Minneapolis, MN 55455, USA}
\address{$^{2}$ Laboratory of Atomic and Solid State Physics, Cornell University, Ithaca, New York 14853, USA}
\address{$^{3}$ Warwick Centre for Predictive Modelling, School of
 Engineering, and Centre for Scientific Computing, University of Warwick, Coventry CV4~7AL, UK}

\begin{abstract}
Fitted interatomic potentials are widely used in atomistic simulations thanks to
their ability to compute the energy and forces on atoms quickly.  However, the simulation results crucially depend on the quality of the potential being used.  Force matching is a method aimed at constructing reliable and transferable interatomic potentials by matching the forces computed by the potential as closely as possible, with those obtained from first principles calculations.  The \potfit program is an implementation of the force-matching method that  optimizes the potential parameters using a global minimization algorithm followed by a local minimization polish.  We extended \potfit in two ways. First, we adapted the code to be compliant with the KIM Application Programming Interface (API) standard (part of the Knowledgebase of Interatomic Models Project). This makes it possible to use \potfit to fit many KIM potential models, not just those prebuilt into the \potfit code. Second, we incorporated the geodesic Levenberg--Marquardt (LM) minimization algorithm into \potfit as a new local minimization algorithm. The extended \potfit was tested by generating a training set using the KIM Environment-Dependent Interatomic Potential (EDIP) model for silicon and using \potfit to recover the potential parameters from different initial guesses.  The results show that EDIP is a ``sloppy model'' in the sense that its predictions are insensitive to some of its parameters, which makes fitting more difficult.  We find that the geodesic LM algorithm is particularly efficient for this case.  The extended \potfit code is the first step in developing a KIM-based fitting framework for interatomic potentials for bulk and two-dimensional materials. The code is available for download via \url{https://www.potfit.net}.
\end{abstract}

\maketitle

\section{Introduction}\label{sec:intro}

An interatomic potential (IP) is a model for approximating the quantum-mechanical interaction of electrons and nuclei in a material through a parameterized functional form that depends only on the positions of the nuclei. IPs such as the Lennard-Jones \cite{jones1924a,jones1924b} and Morse \cite{morse1929} potentials were initially introduced as part of theoretical studies of material behavior in the early 20th century. Interest in IPs renewed in the 1960s following the development of Monte Carlo and molecular dynamics (MD) simulation methods. This led to the development of IPs for a variety of material systems, a trend that accelerated in the 1980s with the formulation of a large number of increasingly complex IPs.  (See \cite{tadmor2011modeling} for more on the history and scientific development of IPs and MD.) Even complex IPs are far less costly to compute than a first principles approach that involves solving the Schr\"{o}dinger equation for the quantum system. This enables simulations of large systems (millions, billions and even trillions of atoms depending on the IP) for very short times or moderate-size systems for longer times (tens or hundreds of nanoseconds). Such simulations can tackle problems that are inaccessible to quantum calculations, such as plastic deformation, fracture, atomic diffusion, and phase transformations \cite{mishin1999interatomic}.

Obtaining an accurate and reliable IP constitutes a challenging problem that involves choosing an appropriate functional form and adjusting the parameters to optimally reproduce a relevant training set of experimental and/or quantum data. Traditionally IPs were fitted to reproduce a set of material properties considered important for a given application, such as the cohesive energy, equilibrium lattice constant, and elastic constants of a given crystal phase. However experience has shown that the \emph{transferability} of such IPs (i.e.\ their ability to accurately predict behavior that they were not fitted to reproduce) can be limited due to the small number of atomic configurations sampled in the training set (although recent work \cite{nichols2016} has shown that this approach can be effective in some cases). Further, as the complexity of IPs increases (both in terms of the functional forms and the number of parameters) it can be difficult to obtain a sufficient number of material properties for the training set. This is particularly true for multispecies systems like intermetallic alloys.

To address these difficulties, Ercolessi and Adams \cite{ercolessi1994interatomic} proposed the \emph{force-matching method} in which a training set containing material properties is augmented with the forces on atoms obtained by first principles calculations for a set of atomic configurations. These can be configurations associated with important structures or simply random snapshots of the crystal as the atoms oscillate at finite temperature. By fitting to this information the transferability of the potential is likely enhanced since it is exposed to a larger cross-section of configuration space. The issue of insufficient training data is also resolved since as many configurations as needed can be easily generated. This makes it possible to increase the number of parameters and in fact in the original Ercolessi--Adams potential for aluminum  \cite{ercolessi1994interatomic} the functional forms were taken to be cubic splines with the spline knots serving as parameters. This gives maximum freedom to the fitting process (at the expense of a clear physical motivation for the functional form). The force-matching method has been widely used since its introduction and there are a number of open source implementations available.  The \emph{ForceFit} \cite{waldher2010forcefit} and \emph{ForceBalance} \cite{wang2012systematic} packages target organic-chemistry applications, while \potfit \cite{brommer2006effective, brommer2007potfit, brommer2015potfit} focuses on solid-state physics.

IP fitting programs (such as those mentioned above) define a ``cost function'' quantifying the difference between the training set data and IP predictions and use global and/or local minimization algorithms to reduce the cost as much as possible.  For example, the \potfit program uses simulated annealing for global minimization followed by a local polish using a conjugate gradient method. A difficulty associated with this procedure is that IPs are nonlinear functions that are often ``sloppy'' in the sense that their predictions are insensitive to certain combinations of their parameters \cite{waterfall:casey:2006}. These soft modes in parameter space can cause the minimization algorithms to fail to converge. Recently an understanding of sloppy models based on ideas from differential geometry has led to efficient methods for fitting such models \cite{transtrum2011geometry} The basic idea is that the parameters of a model (like an IP) define a manifold in the space of data predictions. Fitting the IP then corresponds to finding the point on the manifold closest to the training set. Using these ideas Transtrum et al.\ \cite{transtrum2011geometry} augmented the Levenberg--Marquardt algorithm with a geodesic acceleration adjustment to improve convergence. The new geodesic Levenberg--Marquardt (LM) algorithm is likely to be more efficient and more likely to converge for sloppy model systems than conventional approaches. Geodesic LM has been applied to a variety of applications in physics and biology, but until now not to the IP fitting problem.

Once an IP is developed it must be integrated into an MD code (or other atomic-scale simulation package).  Traditionally this was done on a code-by-code basis. This creates a barrier for the universal usability of IPs.  Users are typically limited by the IPs that happen to be available in the package that they choose.  The same applies to IP fitting programs, which support a limited set of functional forms that they can fit.  Extending codes to support new IPs is a time consuming and error prone process.  It also leads to an inevitable uncontrolled mutation of IPs as modifications to functional forms and/or parameters are introduced (often without adequate documentation).  More recently, emphasis have been placed on archival storage of IPs within projects such as the NIST Interatomic Potential Repository (IPR) \cite{nistpotentials} and the Open Knowledgebase of Interatomic Models (KIM) project \cite{tadmor2011kim,tadmor2013nsf,kimwebsite}.  The NIST IPR stores tabulated parameters for a limited number of functional forms (mostly the embedded-atom method (EAM) \cite{daw1984embedded, daw89model, daw1993embedded}). OpenKIM hosted at \url{https://openkim.org} is a more general effort in which the IP implementation itself is stored in a format that conforms to an application programming interface (API) standard developed as part of the KIM project \cite{kimapi}. The KIM API enables any IP stored in KIM to work seamlessly with any KIM-compliant simulation code including ASAP \cite{asap}, DL\_POLY \cite{smith1996,dlpoly}, GULP \cite{gale1997,gulp}, IMD \cite{stadler1997,roth2013,imd}, LAMMPS \cite{plimpton1995,lammps}, libAtoms+QUIP \cite{quip} and Quasicontinuum \cite{TadmorOrtiz1996}. In addition IPs stored in OpenKIM are exhaustively tested by computing their prediction for a large number of material properties. These calculations are performed by ``KIM Tests'' also archived in OpenKIM. Users can extend the OpenKIM system by uploading their own Tests. The predictions of the IPs can be explored using text-based searches and a user-extendible visualization system \cite{kimviz}.

In this paper we describe a KIM-compliant implementation of the \potfit program designed for fitting sloppy IPs. The work has two main objectives.  First, we extend the \potfit program to support the KIM API making it possible to fit IPs with arbitrary functional forms that are portable to a large number of KIM-compliant simulation codes. Second, we explore the efficiency of the geodesic LM algorithm for fitting IPs using the force-matching method within the \potfit framework. We test the new implementation on a KIM-compliant implementation of the environment-dependent interatomic potential (EDIP) for silicon \cite{bazant1996modeling, bazant1997environment, justo1998edip}. We use the EDIP model to generate a training set of force configurations and explore the ability of \potfit to converge back to the original EDIP parameters from different initial guesses. We find that EDIP has the properties of a sloppy model and that the geodesic LM algorithm can converge to the correct results in situations where other minimizers in \potfit fail.

The paper is organized as follows.  In sections \ref{sec:potfit} and \ref{sec:kim}, we briefly describe the \potfit program and the KIM project, respectively. This is followed by a discussion of the modifications that have been made to \potfit in order to support the KIM API and the geodesic LM algorithm. In \sref{sec:app:edip}, we present the application of the extended \potfit to the EDIP model.  We conclude in \sref{sec:summary} with a summary and a discussion of future directions.

\section{The \potfit program for fitting interatomic potentials}\label{sec:potfit}

The \potfit program\cite{brommer2006effective,brommer2007potfit,brommer2015potfit} (\url{http://www.potfit.net}) is an open source implementation of the force-matching method. It optimizes an IP's parameters by minimizing a cost function defined as the weighted sum of squares of deviations between calculated and reference quantities for forces, energies and/or stresses. (The particular form of the cost function used in this paper is described in \sref{sec:app:edip}.) \potfit consists of two largely independent components: IP implementations, and least-squares minimization.  This program architecture makes it comparatively easy to add either an IP model or a minimization algorithm, which facilitated both aspects of the current work.

\potfit supports both analytic and tabulated IPs.  An analytic IP has a fixed functional form with certain free parameters.  Consider for example the simplest pair potential form where the total energy of $n$ interacting atoms is given by
\[
V = \frac{1}{2} \sum_{\alpha=1}^n \sum_{\beta=1}^n \phi(r_{\alpha\beta}),
\]
in which $r_{\alpha\beta}$ is the distance between atoms $\alpha$ and $\beta$, and
$\phi(r_{\alpha\beta})$ is the energy in the bond connecting them. An example of an analytic function, is the Lennard-Jones potential for which
\[
\phi(r)=4\epsilon \left[
\left(\frac{\sigma}{r}\right)^{12} - \left(\frac{\sigma}{r}\right)^6
\right],
\]
where $\epsilon$ and $\sigma$ are the IP parameters. In contrast, for a tabulated IP, the functional form is arbitrarily defined as an interpolation between sampling points that serve as the IP parameters.  For example, for a cubic spline, given a set of distances $(r_1, r_2, \cdots, r_N)$ and a corresponding set of function values $(\phi_1, \phi_2, \cdots, \phi_N)$, the functional form in the interval $[r_i,r_{i+1}]$ is defined as
\[
\phi(r) = a_i(r-r_i)^3 + b_i(r-r_i)^2+c_i(r-r_i)+d_i,
\]
where the coefficients $(a_i,b_i,c_i,d_i)$ are obtained from the conditions that $\phi(r_i)=\phi_i$ and $\phi(r_{i+1})=\phi_{i+1}$ (with different conditions at the ends that define the type of cubic spline). In this scenario the tabulated function values $(\phi_1, \phi_2, \cdots, \phi_N)$ are the IP parameters.  The form of interpolation used (cubic, quartic, quintic, or otherwise) can have a strong influence on the predictions of the IP \cite{wen2015interpolation}. This is why the interpolation is considered part of the IP by OpenKIM (see \sref{sec:kim}) and stored along with the parameters. The \potfit program supports various functional forms including a variety of pair potentials, embedded atom method (EAM) potentials \cite{daw1984embedded,daw89model,daw1993embedded}, angular-dependent potentials (ADP) \cite{mishin2005phase}, bond-order potentials (Tersoff potential \cite{tersoff1986new,tersoff1989modeling}), and induced dipole potentials (Tangney-Scandolo \cite{Tangney:2002:8898}).

Once an IP is selected, different local and global minimization algorithms are available within \potfit to find the optimal set of parameters that minimize the cost function. The \emph{local} minimization algorithm is Powell's method\cite{Powell:1965:303}, which is a variant of the conjugate gradient (CG) method. Standard CG requires the derivative of the cost function with respect to the IP parameters. The advantage of Powell's method is that it constructs conjugate directions without requiring derivatives. Powell's algorithm begins with an initial guess for the IP parameters and converges to a nearby minimum.  For highly nonlinear IPs it is possible that different initial guesses will lead to different solutions. To address this, \emph{global} minimization algorithms attempt to explore the parameter space to find the deepest minimum. \potfit includes simulated annealing and differential evolution in this class of methods.  See the original \potfit publications\cite{brommer2006effective,brommer2007potfit,brommer2015potfit} for detailed descriptions of the optimization algorithms.

\section{The Open Knowledgebase of Interatomic Models (KIM) project}
\label{sec:kim}

\begin{figure}[b]
\centering
\includegraphics[width=\columnwidth]{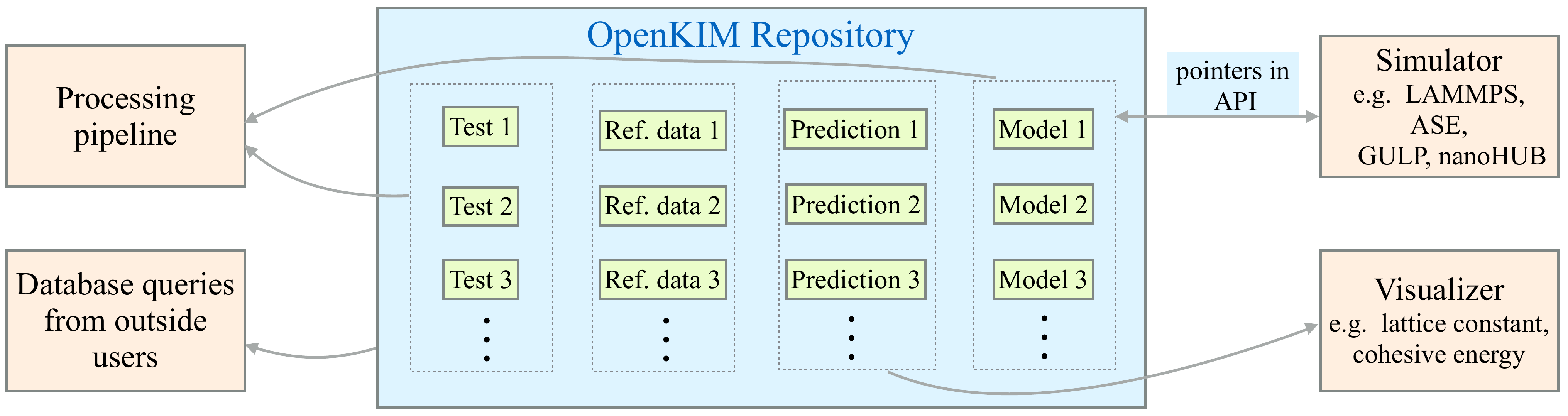}
\caption{OpenKIM framework}
\label{fig:kim:framework}
\end{figure}

The KIM project \cite{tadmor2011kim,tadmor2013nsf,kimwebsite} is an international effort aimed at improving the reliability and accuracy of molecular simulations using IPs.  A schematic of the OpenKIM cyberinfrastructure is displayed in \fref{fig:kim:framework}.  First and foremost OpenKIM provides a repository for IPs at \url{https://openkim.org}. Within the OpenKIM system IPs are referred to more generically as \emph{models}. A KIM Model consists of the computer implementation of the IP along with any parameters. A KIM IP can be a stand-alone model, or a \emph{model driver} that reads in different parameter files to define different models. All content in the OpenKIM system is archived subject to strict provenance control and can be cited using permanent links. For example, later in this article we will be analyzing the EDIP model for silicon \cite{bazant1996modeling,bazant1997environment, justo1998edip} archived in OpenKIM \cite{MO_958932894036_001,MD_506186535567_001,tadmor2011kim}. The citations contain a unique 12-digit KIM identifier and 3-digit version number. This makes it possible to access the actual IP used in this publication at any later date to reproduce the calculations --- an ability lacking prior to OpenKIM archiving. All content within the OpenKIM repository is citeable in this manner and is accessible to external users through web queries.

In addition to archiving IPs, OpenKIM is tasked with determining the suitability of an IP for a given application. Since IPs are approximate, their transferability to phenomena they were not fitted to reproduce is limited. To help define a rigorous method for IP selection, OpenKIM includes a user-extendible suite of \emph{KIM Tests} for characterizing an IP. Each KIM Test computes the prediction of an IP for a well-defined material property (such as the surface free energy for a given crystal face under specified temperature and pressure). Each IP in the system is mated with all compatible KIM Tests using an automated system called the \emph{Processing Pipeline} \cite{bierbaum2016}. Whenever a new Model or Test is uploaded to OpenKIM by members of the community, the pipeline launches to perform the new calculations. The results are stored in the OpenKIM repository and can be explored and compared with first principles and experimental reference data using text-based and graphical visualization tools \cite{kimviz}. Machine learning based tools are being developed to use the accumulated information on the accuracy of an IPs predictions to help select for applications of choice \cite{karls2016thesis}.

The operation of the pipeline involves the coupling of pairs of computer codes, Model and Test, which may be written in different computer languages and need to exchange information. Specifically, the Test repeatedly calls the Model with new input (typically positions and species of atoms) and requires the generated output (typically energy and forces on atoms). This is made possible because all codes in OpenKIM conform to the KIM API \cite{kimapi}. This is a standard developed as part of the KIM project for atomistic simulation codes to exchange information. The KIM API is a lightweight efficient interface where only pointers are exchanged between the interfacing codes, which can be written in any supported language (C, C++, Fortran 77/90/95/2003, Python). A number of major simulation codes currently support the KIM API (listed in \sref{sec:intro}). Anyone of these codes can work seamlessly (without having to be recompiled) with any KIM Model stored on the same computer as a dynamically linked library. (This is akin to a CD being placed in a player from any manufacturer and having its content played.) This greatly increases the portability of IPs. It allows researchers to select the appropriate IP for their application and use it with their code of choice, rather than being limited to the IPs that happen to come with it.

\section{Extensions to \potfit: Geodesic LM and KIM-compliance}
\label{sec:adaption}

An extended version of the \potfit code has been developed that includes a geodesic LM optimizer and support for the KIM API. The new code is available via \url{https://www.potfit.net}. Details of the extensions are given below.

\subsection{Geodesic Levenberg--Marquardt (LM) algorithm}
\label{sec:glm}
As explained in the introduction, IPs often have the characteristics of sloppy models.  This can cause the standard minimization algorithms in \potfit to fail to converge. As an alternative we have incorporated into \potfit the newly developed geodesic LM algorithm \cite{transtrum2010nonlinear,transtrum2011geometry,transtrum2012improvements, transtrum2012geodesic}. Geodesic LM is an efficient method offering a higher likelihood of convergence for sloppy models. The performance of the geodesic LM method is discussed in \sref{sec:perform:min:methods}.  The method is described next.

\smallskip
The least-squares optimization problem is to find the parameters $\boldsymbol{\theta} = \{\theta_1,\ldots,\theta_N\}$ that minimize the cost function $C: \mathbb{R}^N \mapsto \mathbb{R}$:
\begin{equation}\label{eq:cost}
C(\boldsymbol{\theta}) = \frac{1}{2} \sum_{m=1}^M (r_m(\boldsymbol{\theta}))^2
                  = \frac{1}{2} \|\boldsymbol{r}(\boldsymbol{\theta})\|^2
                  = \frac{1}{2} \boldsymbol{r}(\boldsymbol{\theta})^\mathrm{T} \boldsymbol{r}(\boldsymbol{\theta}),
\end{equation}
where the residual $\boldsymbol{r}:\mathbb{R}^N \mapsto \mathbb{R}^M$ is an $M$-dimensional vector function of $N$ parameters. The gradient of $C$ is
\begin{equation}\label{eq:cost:gradient}
\frac{\partial C}{\partial \theta_i}(\boldsymbol{\theta})
    = \sum_{m=1}^M r_m \frac{\partial r_m}{\partial \theta_i}
    \quad\Leftrightarrow\quad
\frac{\partial C}{\partial \boldsymbol{\theta}}(\boldsymbol{\theta})
    = \boldsymbol{J}^\mathrm{T}\boldsymbol{r},
\end{equation}
where $\boldsymbol{J}$ is the Jacobian of $\boldsymbol{r}$, and the Hessian of $C$ is,
\begin{equation}\label{eq:cost:hessian}
\fl\frac{\partial^2 C}{\partial \theta_i \theta_j}(\boldsymbol{\theta})
    = \sum_{m=1}^M \left( \frac{\partial r_m} {\partial \theta_i} \frac{\partial r_m}{\partial \theta_j}
    + r_m \frac{\partial^2 r_m}{\partial \theta_i  \theta_j}  \right)
    \quad\Leftrightarrow\quad
\frac{\partial^2 C}{\partial \boldsymbol{\theta}\partial\boldsymbol{\theta}}(\boldsymbol{\theta})
   = \boldsymbol{J}^\mathrm{T} \boldsymbol{J} + \boldsymbol{r}^T \boldsymbol{K},
\end{equation}
where $\boldsymbol{K}$ is the second derivative of $\boldsymbol{r}$ with respect to $\boldsymbol{\theta}$ (a third-order matrix).

Before introducing geodesic LM, let us consider the more basic Gauss--Newton method. This approach is based on a linear approximation of $\boldsymbol{r}$\cite{madsen2004methods},
\begin{equation}\label{eq:lin:approx}
\boldsymbol{r}(\boldsymbol{\theta}+\delta\boldsymbol{\theta})  \approx \boldsymbol{r}(\boldsymbol{\theta}) + \boldsymbol{J}(\boldsymbol{\theta})\delta\boldsymbol{\theta}.
\end{equation}
Substituting \eref{eq:lin:approx} into \eref{eq:cost} we see that,
\begin{eqnarray}\label{eq:cost:approx}
C(\boldsymbol{\theta} + \delta\boldsymbol{\theta})
    &= \frac{1}{2} \boldsymbol{r}(\boldsymbol{\theta} + \delta\boldsymbol{\theta} )^\mathrm{T} \boldsymbol{r}(\boldsymbol{\theta}  + \delta\boldsymbol{\theta}) \nonumber\\
    &= \frac{1}{2} (\boldsymbol{r} + \boldsymbol{J}\delta\boldsymbol{\theta})^\mathrm{T}(\boldsymbol{r} + \boldsymbol{J}\delta\boldsymbol{\theta}) \nonumber\\
    &= C(\boldsymbol{\theta})
+ \delta\boldsymbol{\theta}^\mathrm{T} \boldsymbol{J}^\mathrm{T} \boldsymbol{r} + \frac{1}{2} \delta\boldsymbol{\theta}^\mathrm{T} \boldsymbol{J}^\mathrm{T} \boldsymbol{J} \delta\boldsymbol{\theta}.
\end{eqnarray}
Taking the first derivative of \eref{eq:cost:approx} with respect to $\delta\boldsymbol{\theta}$ and setting it to zero, we have,
\begin{equation}\label{eq:gauss:newton}
\delta\boldsymbol{\theta}
    = - (\boldsymbol{J}^\mathrm{T} \boldsymbol{J})^{-1} \boldsymbol{J}^\mathrm{T}\boldsymbol{r}.
\end{equation}
In \eref{eq:gauss:newton} $\delta\boldsymbol{\theta}$ is a local decent direction of $C$, thus we can obtain the local minimum of $C$ by solving \eref{eq:gauss:newton} iteratively. This is the Gauss--Newton method.

Levenberg \cite{levenberg1944algorithm} suggested a ``damped version'' of \eref{eq:gauss:newton} where the parameters are iteratively updated according to
\begin{equation}\label{eq:lm:del:theta}
\delta\boldsymbol{\theta}
    = - (\boldsymbol{J}^\mathrm{T} \boldsymbol{J} +\lambda\boldsymbol{D}  )^{-1} \boldsymbol{J}^\mathrm{T}\boldsymbol{r}, \quad{\lambda\geq 0}.
\end{equation}
Here $\lambda$ is a (non-negative) damping parameter, and $\boldsymbol{D} =  \boldsymbol{I}$ is the identity.  The Levenberg method is an interpolation between the Gauss--Newton algorithm and the steepest decent algorithm. For small $\lambda$, \eref{eq:lm:del:theta} reduces to the Gauss--Newton equation in \eref{eq:gauss:newton}; whereas for large $\lambda$, $\delta\boldsymbol{\theta} \approx -\lambda^{-1}\boldsymbol{J}^\mathrm{T}\boldsymbol{r}$, which lies along the gradient (i.e.\ the steepest descent direction) in \eref{eq:cost:gradient}.

In order to overcome slow convergence in directions of small gradients, Marquardt \cite{marquardt1963algorithm} proposed taking $\boldsymbol{D} = \mathrm{diag}(\boldsymbol{J}^\mathrm{T} \boldsymbol{J})$, where $\mathrm{diag}(\boldsymbol{A})$ returns a diagonal matrix with elements equal to the diagonal elements of the square matrix $\boldsymbol{A}$). The downside of this approach is that it greatly increases the susceptibility for parameter evaporation \cite{transtrum2012improvements}, i.e.\ the algorithm pushes the parameters to infinite values without finding a good fit around saddle points.

To improve the efficiency of the Levenberg--Marquardt (LM) method, Transtrum and Sethna \cite{transtrum2010nonlinear, transtrum2011geometry, transtrum2012improvements, transtrum2012geodesic} proposed the geodesic LM algorithm in which the minimization step size is constrained and the residual is modified to include a second-order (geodesic acceleration) correction,
\begin{equation}
\boldsymbol{r}(\boldsymbol{\theta}+\delta\boldsymbol{\theta})
\approx \boldsymbol{r}(\boldsymbol{\theta})
+ \boldsymbol{J}(\boldsymbol{\theta})\delta\boldsymbol{\theta}
+ \frac{1}{2} \delta\boldsymbol{\theta}^\mathrm{T} \boldsymbol{K}(\boldsymbol{\theta}) \delta\boldsymbol{\theta}.
\end{equation}
The expansion of the cost function is then
\begin{equation}\label{eq:geodesic:cost}
C(\boldsymbol{\theta} + \delta\boldsymbol{\theta})
= \frac{1}{2} \boldsymbol{r}(\boldsymbol{\theta} + \delta\boldsymbol{\theta} )^\mathrm{T} \boldsymbol{r}(\boldsymbol{\theta} + \delta\boldsymbol{\theta})
+ \frac{1}{2}\lambda \delta\boldsymbol{\theta}^\mathrm{T} \boldsymbol{D}  \delta\boldsymbol{\theta},
\end{equation}
where a Lagrange multiplier enforces the constraint on the step size $\delta\boldsymbol{\theta}$. Taking the first derivative of \eref{eq:geodesic:cost} with respect to $\delta\boldsymbol{\theta}$ and setting it to zero, we have
\begin{equation}\label{eq:geodesic:cost:deriv}
\fl J_{mi}r_m + (J_{mi}J_{mj} + r_mK_{mij} + \lambda D_{ij})\delta\theta_j
+ (J_{mk}K_{mij} +\frac{1}{2} J_{mi}K_{mkj})\delta\theta_j \delta\theta_k
= 0,
\end{equation}
where the indices are included explicitly to avoid ambiguity and Einstein's summation convention over repeated indices is adopted.  The step size $\delta\theta_i$ is written as a sum of two parts:
\begin{equation}\label{eq:del1:del2}
\delta\theta_i = \delta\theta^{(1)}_i + \delta\theta^{(2)}_i,
\end{equation}
where $\delta\theta^{(1)}_i$ is given by \eref{eq:lm:del:theta} and $\delta\theta^{(2)}_i$ includes the remaining terms.  Substituting \eref{eq:del1:del2} and \eref{eq:lm:del:theta} into \eref{eq:geodesic:cost:deriv}, $\delta\theta^{(2)}_i$ can be solved as,
\begin{eqnarray}\label{eq:geodesic:deltheta2}
\delta\theta^{(2)}_i
&= -\frac{1}{2}(J_{mi}J_{mj} + \lambda D_{ij})^{-1} J_{pj} K_{pkl} \delta\theta^{(1)}_k\delta\theta^{(1)}_l \nonumber\\
&= -\frac{1}{2}(J_{mi}J_{mj} + \lambda D_{ij})^{-1} r_p'',
\end{eqnarray}
where some small terms are ignored (see \cite{transtrum2012improvements} for details) and the directional second derivative $ r_p'' = K_{pkl} \delta\theta^{(1)}_k\delta\theta^{(1)}_l $ is used.

As seen from \eref{eq:geodesic:deltheta2}, the geodesic acceleration correction only depends on the directional second derivative oriented along the first order correction $\delta\boldsymbol{\theta}^{(1)}$.  This feature is very important because the cost to compute the directional second derivative is reasonably small and hence will not overly add to the computational burden.

Finally, in order to prevent parameter evaporation in the geodesic LM algorithm and to increase the likelihood of convergence, allowable step sizes must satisfy:
\begin{equation}
\frac{2 \| \delta\boldsymbol{\theta}^{(2)} \|}{\| \delta\boldsymbol{\theta}^{(1)} \|} \leq \alpha,
\end{equation}
where $\alpha$ is some parameter (usually smaller than 1) and whose optimal value depends on the specific problem.

An open source implementation of the geodesic LM algorithm has been made available by Transtrum \cite{geodesicLM}. In addition to the geodesic acceleration correction, the geodesic LM package includes different options for updates to the damping parameter and the damping matrix, different conditions for accepting uphill steps (i.e.\ steps that increase $C$), and whether or not to use Broyden's method \cite{broyden1965class} to lower the Jacobian update frequency.  In this work, the damping matrix is set to $\boldsymbol{I}$, Umrigar and Nightingale's method\cite{transtrum2012improvements} is used to update the damping parameter and to decide whether to accept uphill steps, and we do not employ Broyden's method to update the Jacobian because the computational cost in this problem is not too high.

\subsection{KIM-compliance}
The original version of \potfit is limited to using IPs built into the program. The philosophy underlying the KIM API is different. It is based on the idea of separation between the simulation code (called a ``simulator'' in KIM parlance) and the IP (called the KIM Model). Whenever the simulator (\potfit in this case) requires the energy or forces for an atomic configuration, it calls a KIM Model with the necessary input (e.g.\ coordinates and species of the atoms) and receives back the required output (e.g.\ the forces on the atoms). Enforcing this separation makes it possible for the simulator to work with any compatible KIM Model in plug-and-play fashion.\footnote{A KIM Model is compatible with a simulator if it supports the required species and boundary conditions that the simulator requires. This is automatically determined by the KIM API using metadata in ``KIM descriptor'' files that accompany the simulator and KIM Model.} The KIM API is the standard for the exchange of information between the communicating codes. In practice this works by having all data exchanged between the simulator and IP stored in a data structure called the ``KIM API Object.'' The simulator and IP can access and change this data using KIM API library routines. The only direct data transfer required between the programs is for the simulator to pass to the KIM Model the pointer to the location in memory of the KIM API Object. This makes the KIM API extremely efficient and provides it with cross language compatibility (as described in \sref{sec:kim}).  This means that \potfit, which is written in C, can work efficiently and seamlessly with KIM Models written in C, C++, Fortran 77/90/95/2003, and Python.

The \potfit program has an unusual requirement for a simulator in that it needs to modify the parameters of the IP that it is optimizing. This is supported by the KIM API through the ability of KIM Models to ``publish their parameters.'' A KIM Model that does so has in its KIM descriptor file a list of its parameters identified by name with the prefix ``PARAM\_FREE\_''. The \potfit user can select any of these parameters to optimize. This is specified in the input file to \potfit along with the initial guesses for the parameters, and the KIM Model identifier (see \sref{sec:kim}).

\begin{figure}
\centering
\includegraphics[width=0.7\columnwidth]{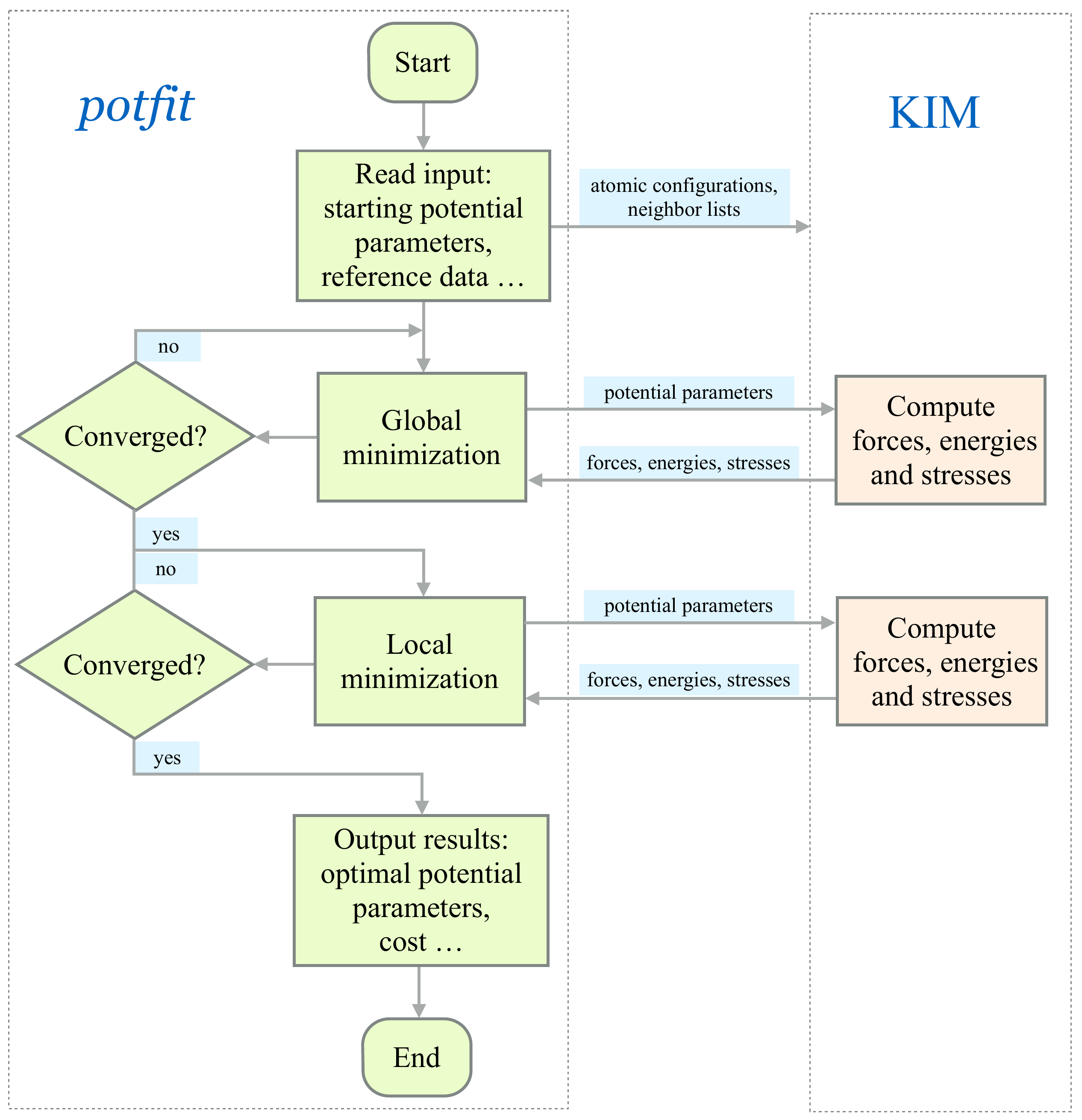}
\caption{Flowchart of the KIM-compliant \potfit program}
\label{fig:potfit:kim}
\end{figure}
The operation of the KIM-compliant implementation of \potfit is described in \fref{fig:potfit:kim}. The program reads in the input defining the optimization problem and then generates one or more KIM API Objects containing the training set to be used for the optimization. At each step in the optimization whenever the IP parameters need to be changed, the following steps are performed using KIM API library calls:
\begin{enumerate}
\item The KIM Model parameters are changed in the KIM API Object;
\item The KIM Model is called to compute the new energy and forces;
\item \potfit acquires the energy and forces from the KIM API Object.
\end{enumerate}
When the optimization is complete, \potfit outputs the final set of parameters and writes out a new KIM Model, which can be uploaded to \url{https://openkim.org} if desired.

\section{Application of \kimpotfit to EDIP}
\label{sec:app:edip}
In this section we test the performance of \kimpotfit by applying it to the optimization of the EDIP model for silicon \cite{bazant1996modeling,bazant1997environment, justo1998edip} archived in OpenKIM \cite{MO_958932894036_001,MD_506186535567_001,tadmor2011kim}. The training set consists of the energy and forces computed with the original EDIP model for a periodic configuration of $5\times 5\times 5$ conventional unit cells for a total of 1000 atoms perturbed off the ideal diamond structure positions.  Our objective is to see whether starting from different initial guesses for the parameters, the optimization procedure is able to recover the EDIP parameters.

\subsection{Cost function}\label{sec:cost}
As explained in \sref{sec:potfit}, IP optimization corresponds to minimizing a cost function quantifying the difference between the IP's predictions and a training set of reference data. Our training set consists of a single atomic configuration with $n=1000$ atoms.  The cost function used in our analysis is
\begin{equation}
C(\boldsymbol{\theta}) = \frac{1}{2}w_1 \sum_{\alpha=1}^n \|\boldsymbol{f}_\alpha -\boldsymbol{f}_{\alpha}^0 \|^2 + \frac{1}{2}w_2 (E-E^0)^2,
\end{equation}
where  $\boldsymbol{f}_\alpha  = \boldsymbol{f}_\alpha(\boldsymbol{\theta})$  and $\boldsymbol{f}_\alpha^0$ are the predicted and reference force on atom $\alpha$, $E=E(\boldsymbol{\theta})$ and $E^0$ are the predicted and reference energy, and $w_1$ and $w_2$ are weights. In our example we take $w_1 =1 \, (\text{\AA/eV})^2$, $w_2 =1 \, (\text{1/eV})^2$ to give a unitless cost function. The reference forces and energy are defined as $\boldsymbol{f}_\alpha^0  = \boldsymbol{f}_\alpha (\boldsymbol{\theta}^0)$ and $E^0=E(\boldsymbol{\theta}^0)$, where $\boldsymbol{\theta}^0$ are the original EDIP parameters. EDIP's functional form and parameters along with the training set used are given in the supplementary information accompanying this article.

The Hessian of the cost function at the original parameters $\boldsymbol{\theta}^0$ follows from \eref{eq:cost:hessian} as
\begin{equation}\label{eq:hessian}
H^0_{ij}=
\left. \frac{\partial ^2C}{\partial \theta_i\partial \theta_j} \right|_{\boldsymbol{\theta}^0}
    = w_1 \sum_{\alpha=1}^M  \left. \frac{\partial \boldsymbol{f}_\alpha} {\partial \theta_i} \right|_{\boldsymbol{\theta}^0} \cdot  \left. \frac{\partial \boldsymbol{f}_\alpha}{\partial \theta_j} \right|_{\boldsymbol{\theta}^0}
    + w_2 \left. \frac{\partial E}{\partial \theta_i}\right|_{\boldsymbol{\theta}^0} \left. \frac{\partial E}{\partial \theta_j} \right|_{\boldsymbol{\theta}^0} ,
\end{equation}
where we have used $\left. (\boldsymbol{f}_\alpha - \boldsymbol{f}_\alpha^0)  \right|_{\boldsymbol{\theta}^0} = 0$ and $\left. (E-E^0)\right|_{\boldsymbol{\theta}^0} = 0$. We will use the Hessian in the next section to evaluate the sensitivity of EDIP predictions to its parameters.

\subsection{EDIP sensitivity analysis}
\label{sec:sensitivity:param}
\begin{figure}
\centering
\includegraphics[width=0.5\columnwidth]{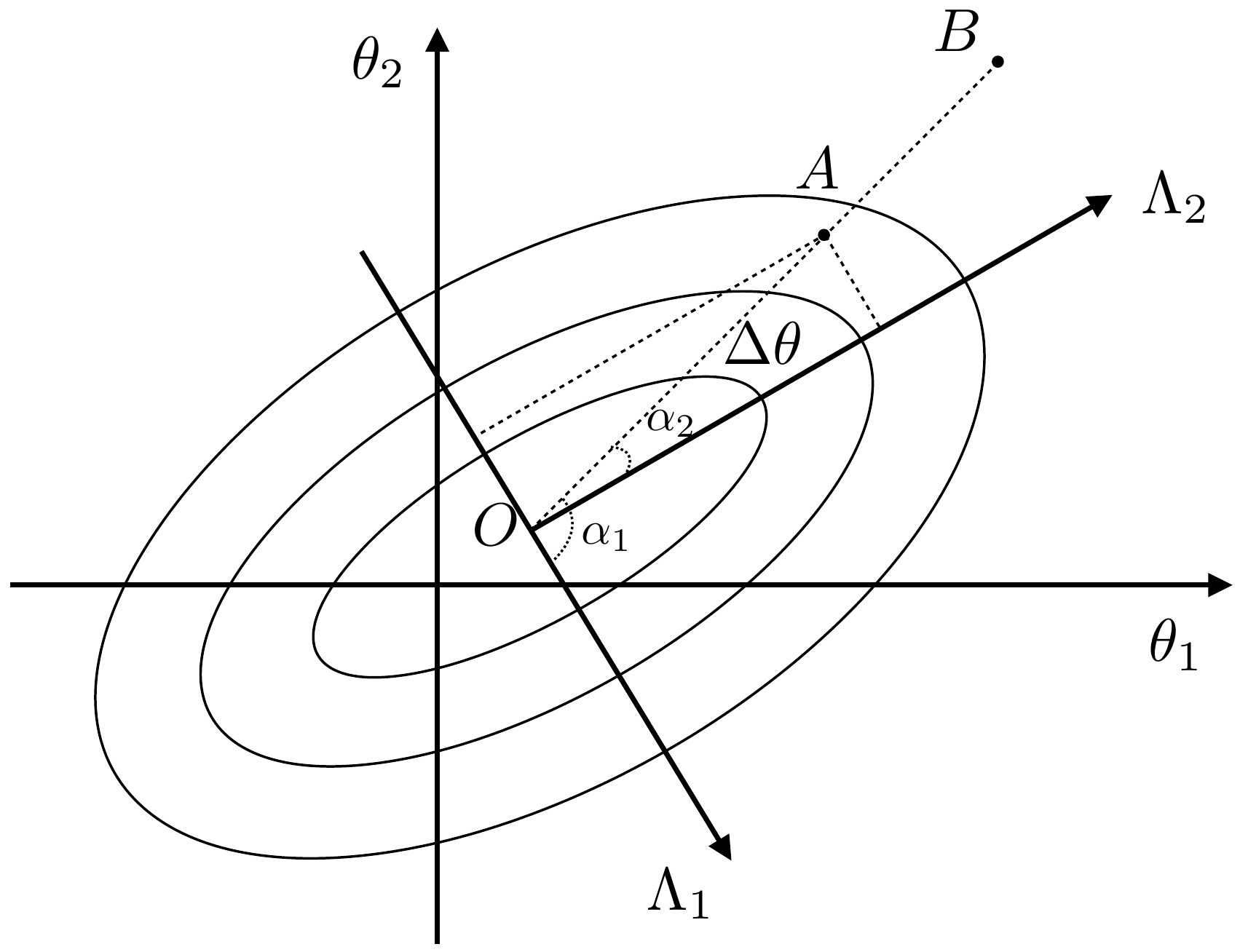}
\caption{Contour plot of a cost function with two parameters. $\Lambda_1$ and $\Lambda_2$ are the eigendirections of the Hessian of the cost function corresponding to the larger and smaller eigenvalues, respectively. Points $A$ and $B$ represent two fits.
}
\label{fig:cost:contour}
\end{figure}

It is a common feature of IPs that the prediction of a model (including the cost function) are weakly dependent on certain directions in parameter space, i.e.\ the models are \emph{sloppy} \cite{frederiksen2004bayesian}. This is demonstrated schematically in \fref{fig:cost:contour} showing the contour plot of a cost function with only two parameters. Clearly the function varies more slowly along the $\boldsymbol{\Lambda}_2$ eigendirection than along $\boldsymbol{\Lambda}_1$. As a result of this structure of $C(\boldsymbol{\theta})$, convergence along certain directions can be incomplete. For example in the figure, the minimum is located at $O$ but a convergence could terminate at a set of parameters offset to the optimal ones by $\Delta\boldsymbol{\theta}$. An indication of whether this is occurring can be obtained by computing the angle between $\Delta\boldsymbol{\theta}$ and the eigenvectors of the Hessian $\partial^2 C/\partial \theta_i\partial \theta_j$ evaluated at $O$,
\begin{equation}\label{eq:cos:alpha}
\cos\alpha_i
= \frac{ \Delta\boldsymbol{\theta} \cdot \boldsymbol{\Lambda}_i} {\|\Delta\boldsymbol{\theta}\|}.
\end{equation}
Point $A$ and point $B$ (corresponding to another fit) are equally close to the eigendirections because they form the same angles.

\begin{figure}
\centering
\includegraphics[width=0.5\columnwidth]{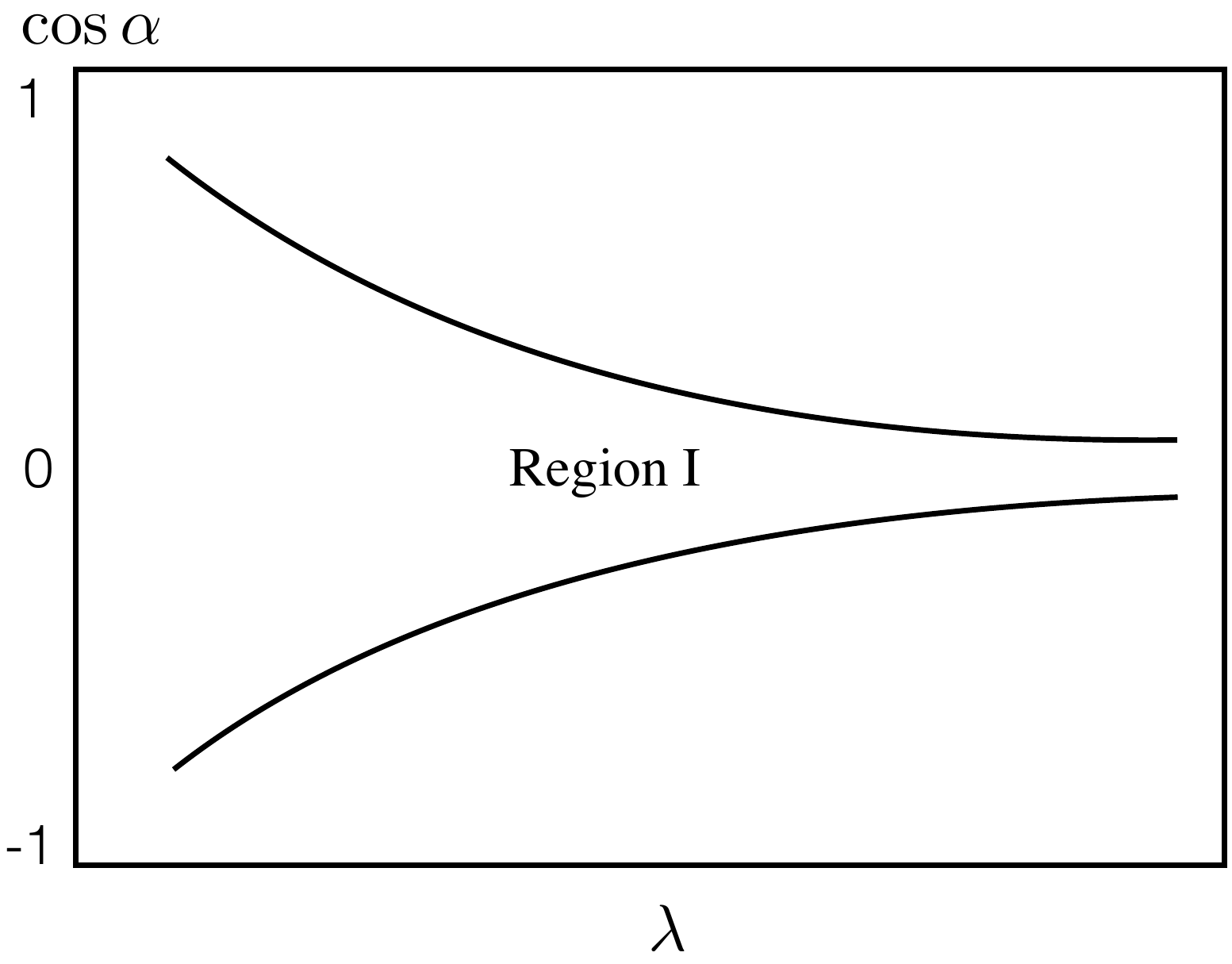}
\caption{Expected region for normalized projected parameter errors if the fits are aligned along the sloppy direction(s). The horizontal axis corresponds to the eigenvalues of the Hessian sorted in increasing order.}
\label{fig:err:region}
\end{figure}

In general, the eigendirections associated with smaller eigenvalues are more ``sloppy'' and we would expect larger scatter in the $\cos\alpha_i$ values computed in \eref{eq:cos:alpha} for an ensemble of fits.  Thus qualitatively we expect a behavior similar to that shown in \fref{fig:err:region}, where the scatter in $\cos\alpha_i$ obtained from a large number of fitting attempts with different initial guesses would occupy a region similar to Region~I in the figure with larger scatter along directions associated with smaller eigenvalues. This would be the signature of a sloppy model.

\begin{figure}
\centering
\begin{subfigure}{0.49\columnwidth}
\includegraphics[width=\columnwidth]{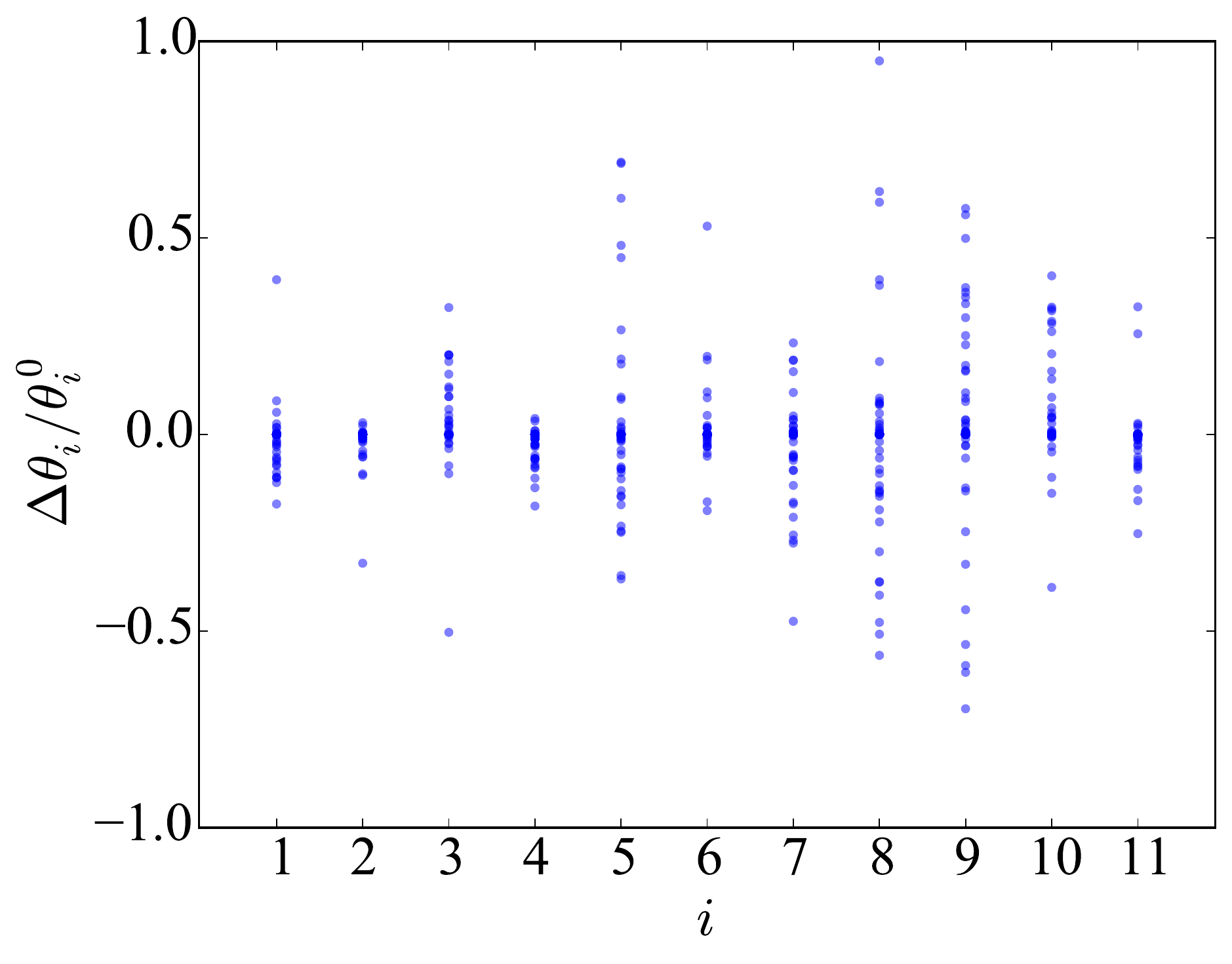}
\caption{}
\label{fig:param_error_a}
\end{subfigure}
\begin{subfigure}{0.49\columnwidth}
\includegraphics[width=\columnwidth]{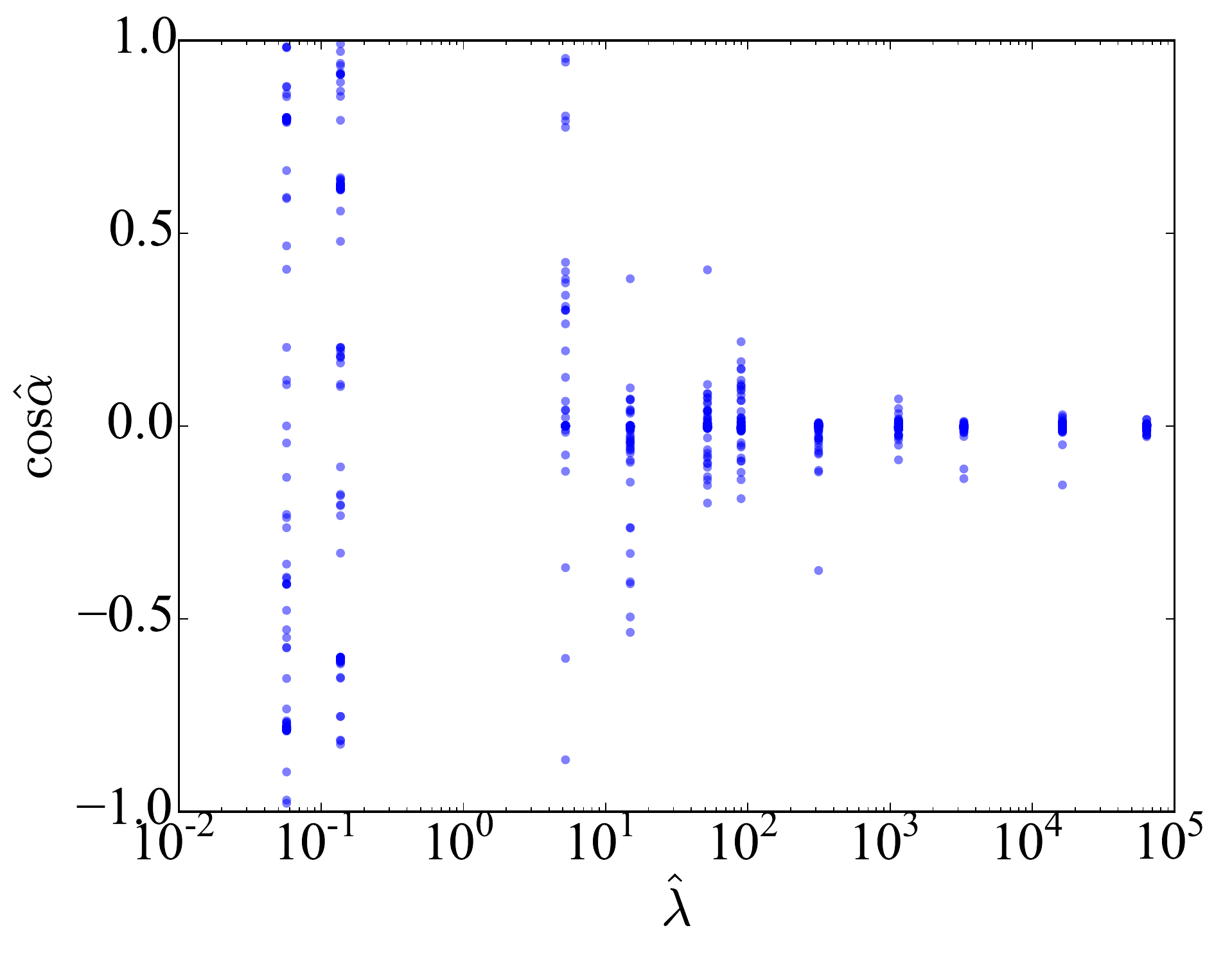}
\caption{}
\label{fig:param_error_b}
\end{subfigure}
\caption{Errors in fitting the EDIP parameters. (a) Relative errors. (b) Normalized error projected on the eigendirections of the Hessian of the cost function in the logarithm parameter space. }
\label{fig:param_error}
\end{figure}

We perform this analysis for EDIP. EDIP has 13 parameters, but only 11 of them are used in this study with the two cutoffs excluded. The training set is described in \sref{sec:cost}. We take an ensemble of 100 initial guesses obtained by adding to the original EDIP parameters random numbers $\rho$ drawn from a Gaussian distribution with average zero and a standard deviation of 0.15, i.e.\ $\theta_i := \theta^0_i\times(1+\rho), i = 1,\ldots,11$, and perform the fitting procedure for each one.  The relative errors $\Delta\theta_i/\theta^0_i=(\theta^*_i-\theta^0_i)/\theta^0_i$ (for $i=1,\ldots,11$) between the parameters $\boldsymbol{\theta}^*$ obtained in the fitting procedure and the original EDIP parameters $\boldsymbol{\theta}^0$ for all 100 optimizations are shown in \fref{fig:param_error_a}.  No clear trend is apparent.  Instead let us plot the cosine defined in \eref{eq:cos:alpha} as done schematically in \fref{fig:err:region}. To improve the scaling of the plot let us modify the definition to use the logarithm of the parameters, $\hat\theta_i = \log\theta_i$. (This is important when different parameters have very different magnitudes.) The definition equivalent to \eref{eq:cos:alpha} is then
\begin{equation}\label{eq:cos:hat:alpha}
\cos\hat\alpha_i
= \frac{\Delta\boldsymbol{\hat\theta} \cdot \boldsymbol{\hat\Lambda}_i} {\|\Delta\boldsymbol{\hat\theta}\|},
\end{equation}
where $\Delta\boldsymbol{\hat\theta}=\boldsymbol{\hat\theta}^*-\boldsymbol{\hat\theta}^0$ and $\boldsymbol{\hat\Lambda_i}$ is the $i$th eigendirection of the Hessian of the cost function $C$ in the logarithm parameter space, $\partial^2 C/\partial \hat\theta_i\partial \hat\theta_j = \theta_i \theta_j \partial^2 C/\partial \theta_i\partial \theta_j$. The results for EDIP are plotted in \fref{fig:param_error_b}, and are very much in line with the expected behavior in \fref{fig:err:region}. In particular, we see that the spread is mostly accounted for by the first four eigenterms each of which corresponds to a particular combination of parameters as quantified by the eigendirections.

%

\subsection{Cost along eigendirections}

\begin{figure}
\centering
\begin{subfigure}{0.32\columnwidth}
\includegraphics[width=\columnwidth]{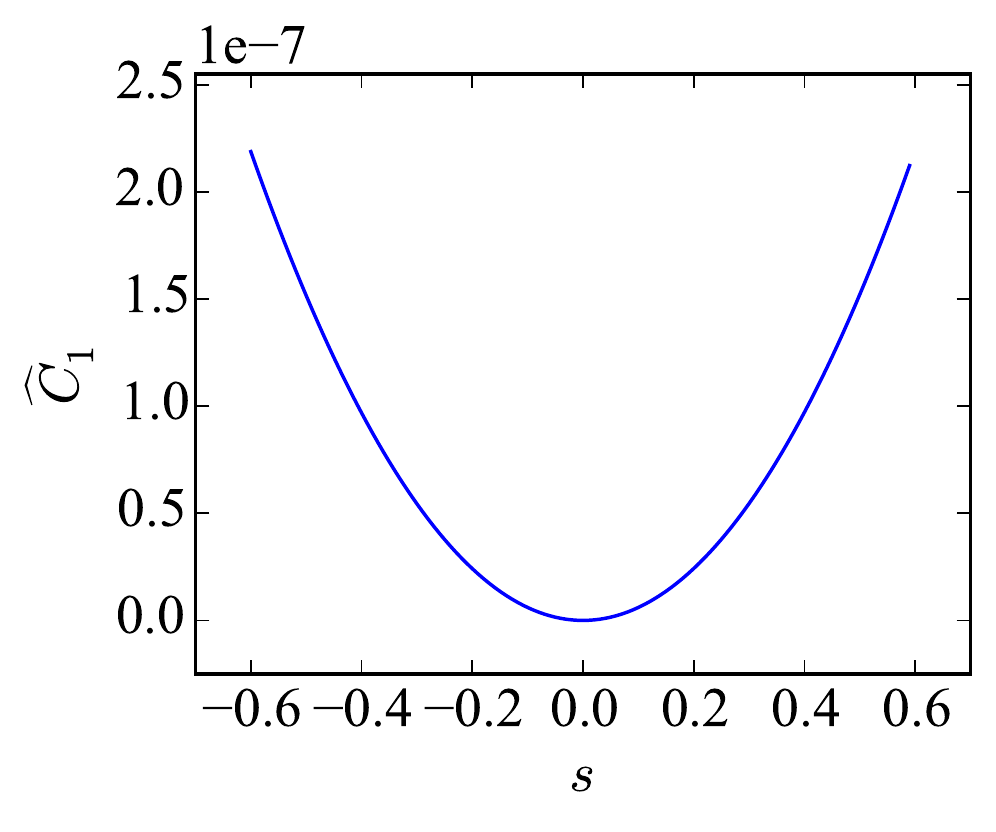}
\end{subfigure}
\begin{subfigure}{0.32\columnwidth}
\includegraphics[width=\columnwidth]{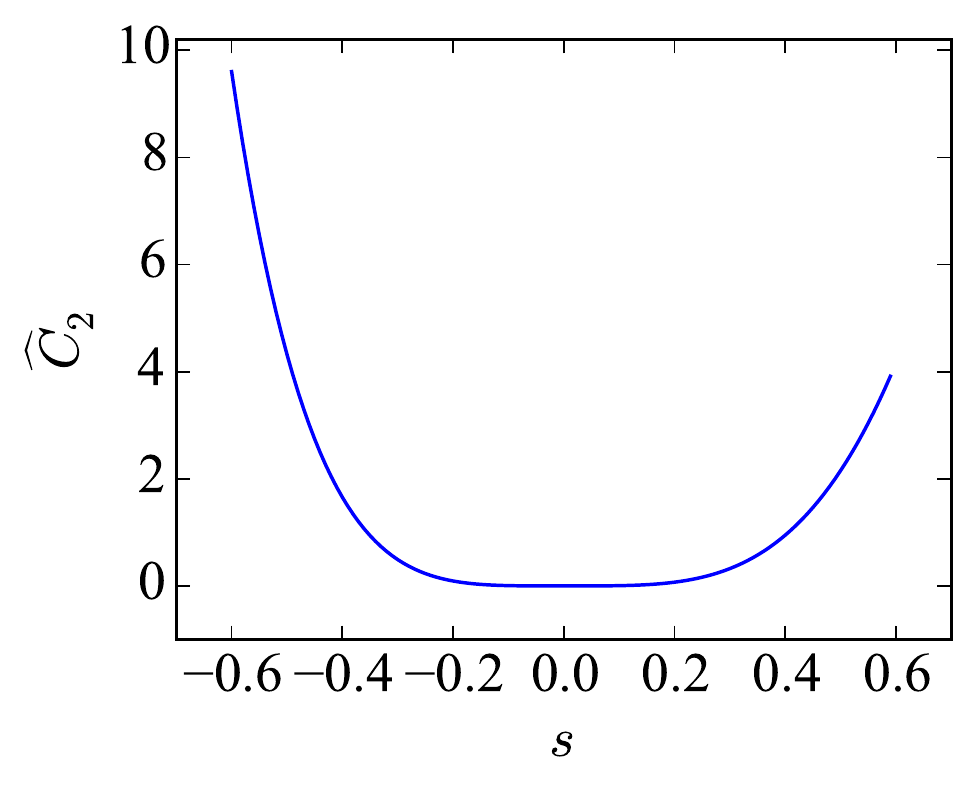}
\end{subfigure}
\begin{subfigure}{0.32\columnwidth}
\includegraphics[width=\columnwidth]{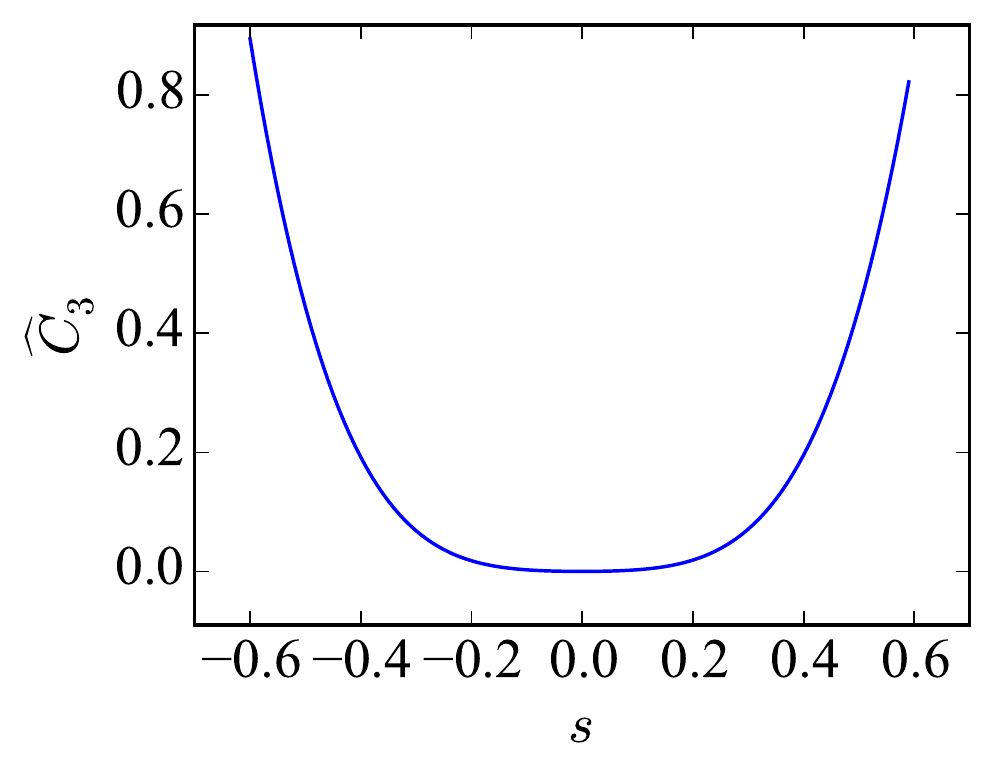}
\end{subfigure}
\begin{subfigure}{0.32\columnwidth}
\includegraphics[width=\columnwidth]{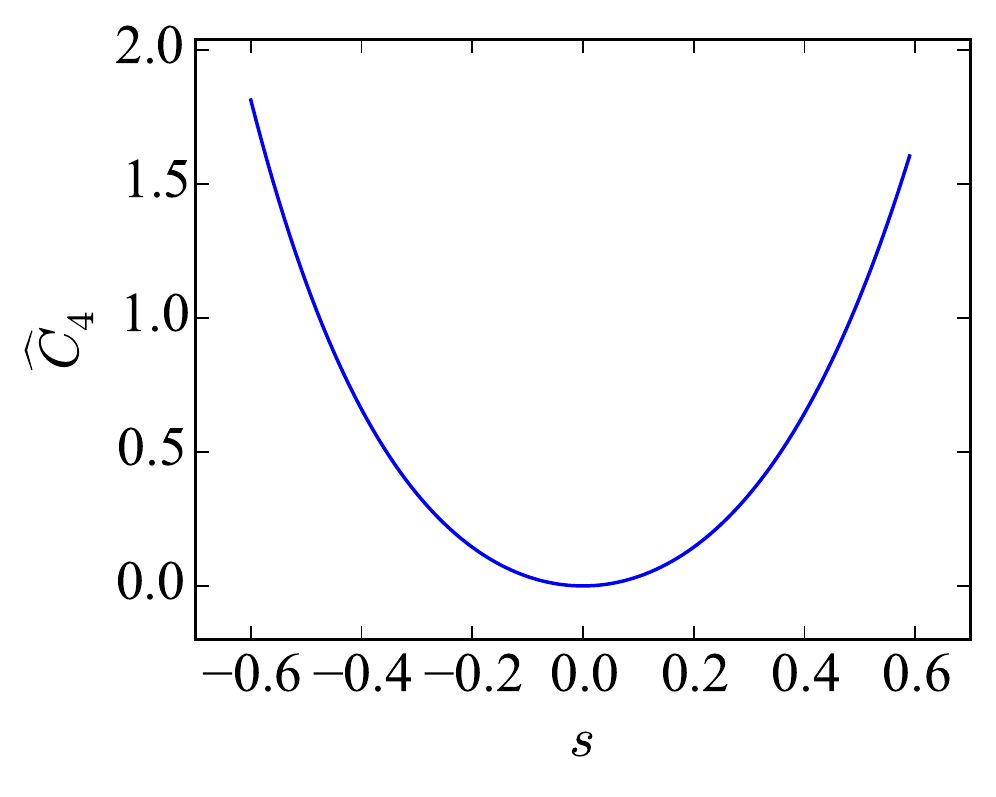}
\end{subfigure}
\begin{subfigure}{0.32\columnwidth}
\includegraphics[width=\columnwidth]{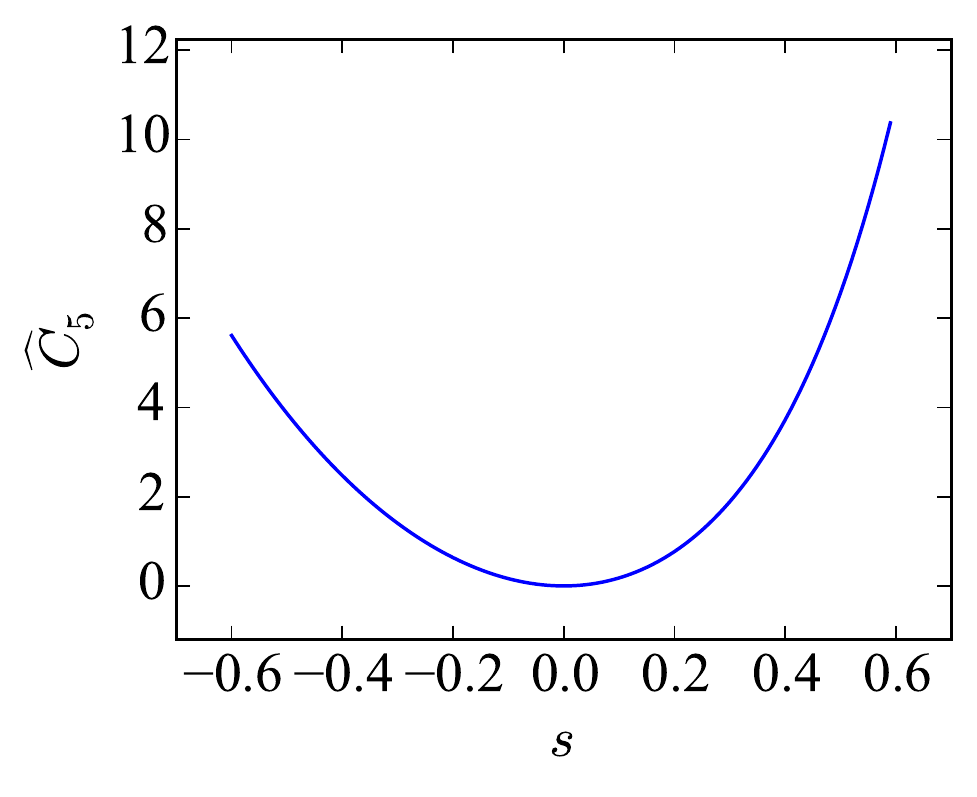}
\end{subfigure}
\begin{subfigure}{0.32\columnwidth}
\includegraphics[width=\columnwidth]{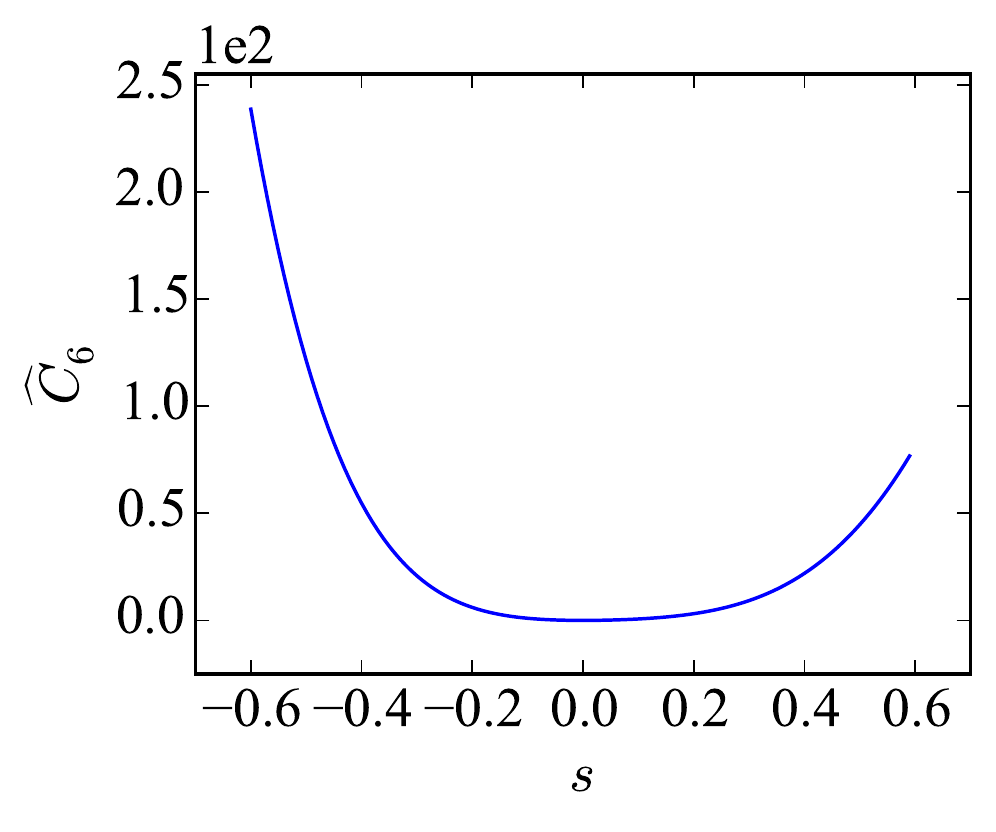}
\end{subfigure}
\begin{subfigure}{0.32\columnwidth}
\includegraphics[width=\columnwidth]{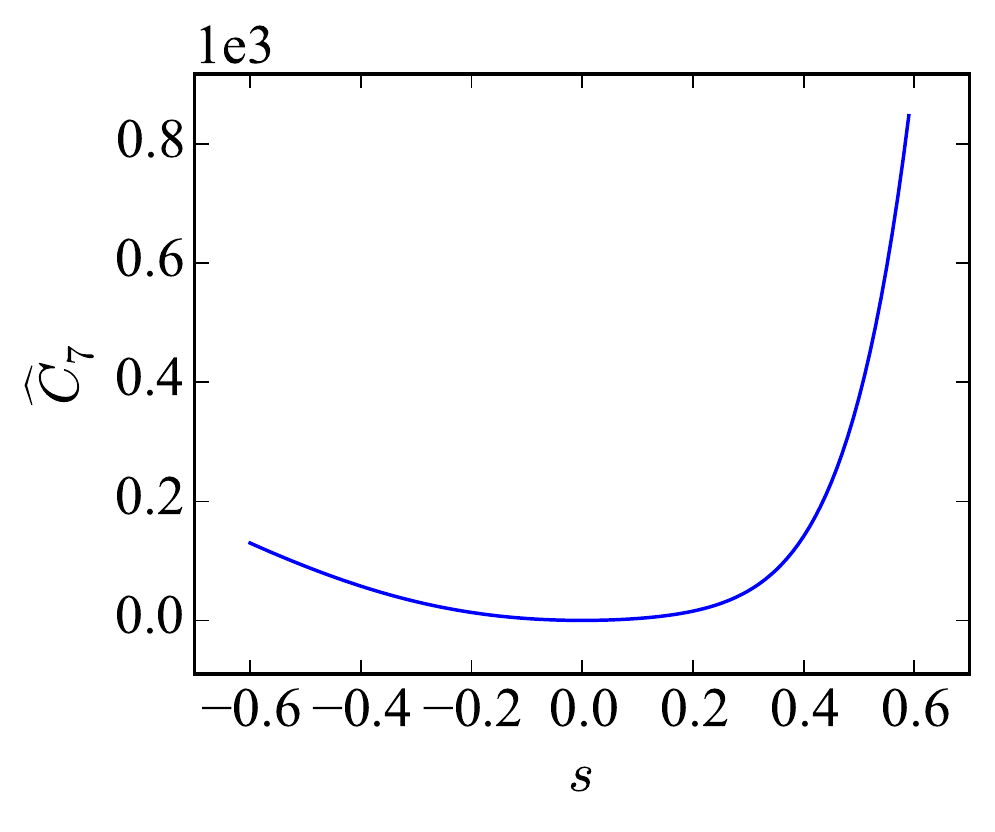}
\end{subfigure}
\begin{subfigure}{0.32\columnwidth}
\includegraphics[width=\columnwidth]{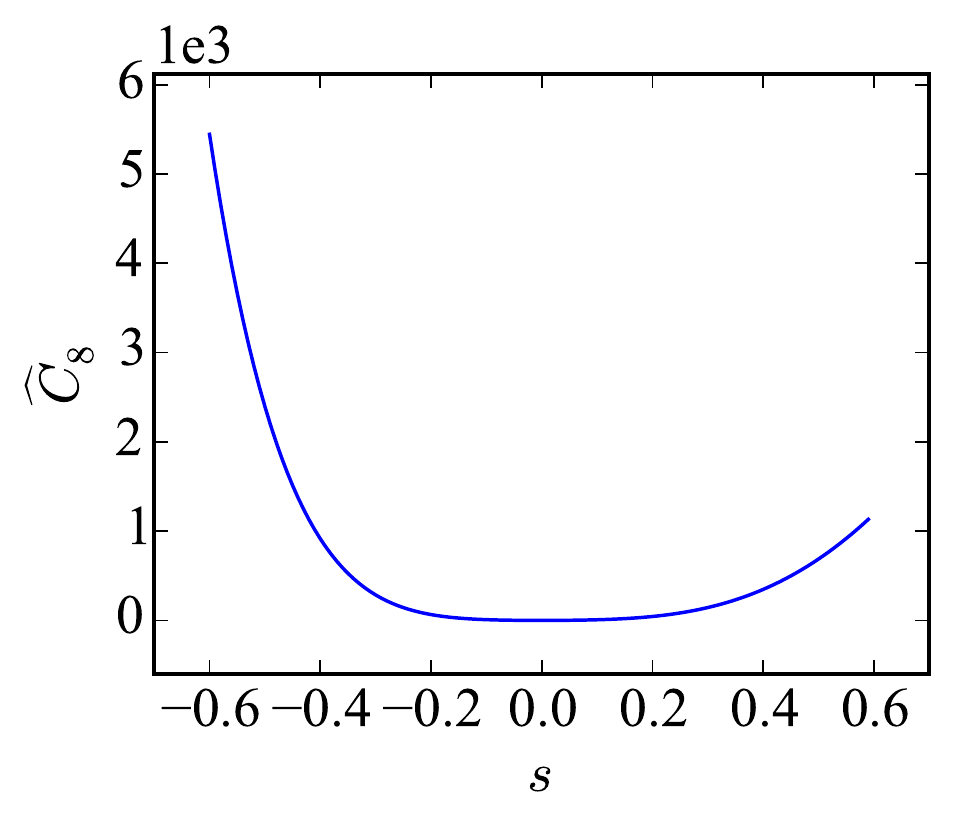}
\end{subfigure}
\begin{subfigure}{0.32\columnwidth}
\includegraphics[width=\columnwidth]{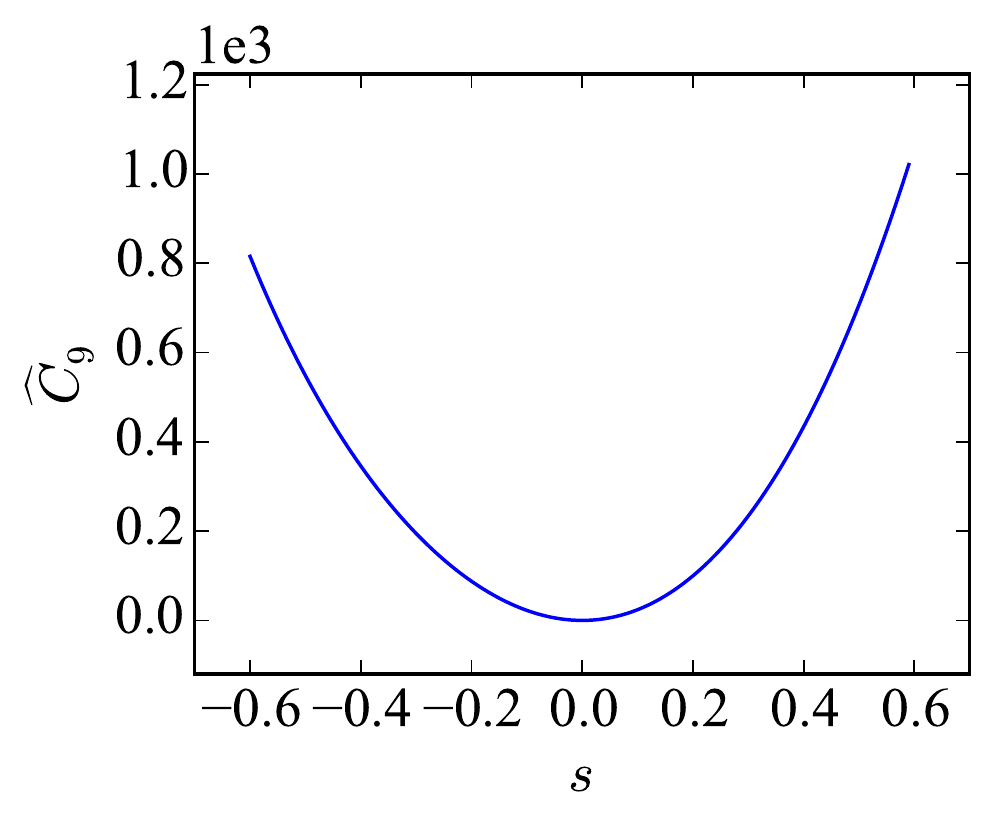}
\end{subfigure}
\begin{subfigure}{0.32\columnwidth}
\includegraphics[width=\columnwidth]{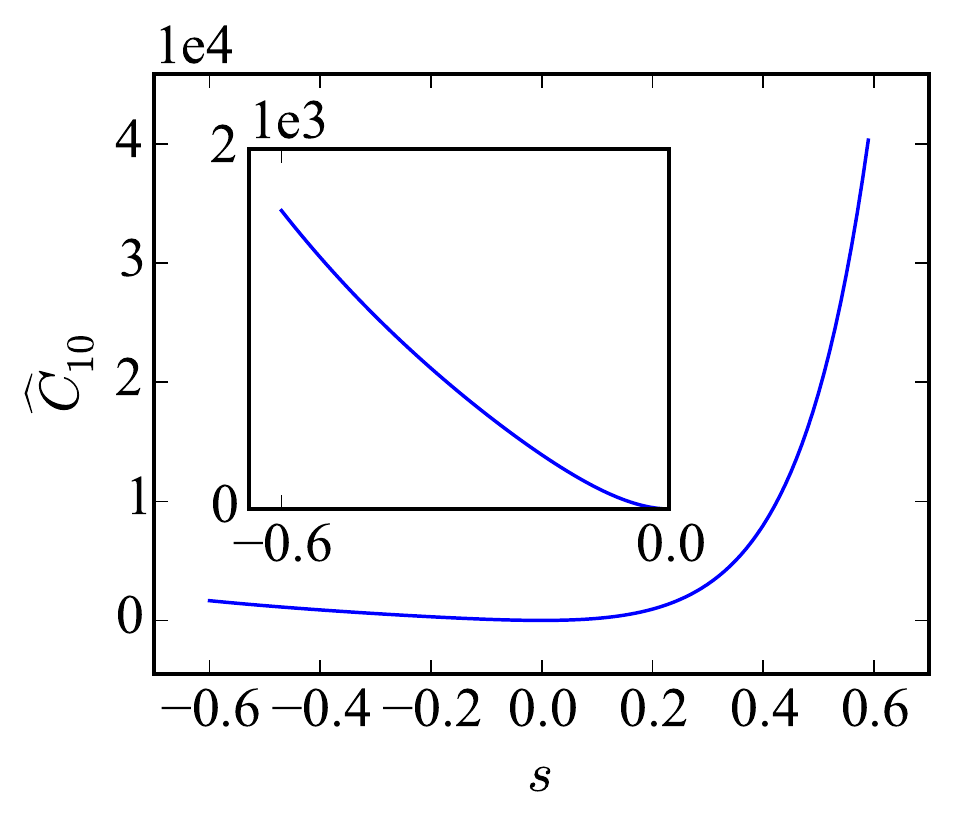}
\end{subfigure}
\begin{subfigure}{0.32\columnwidth}
\includegraphics[width=\columnwidth]{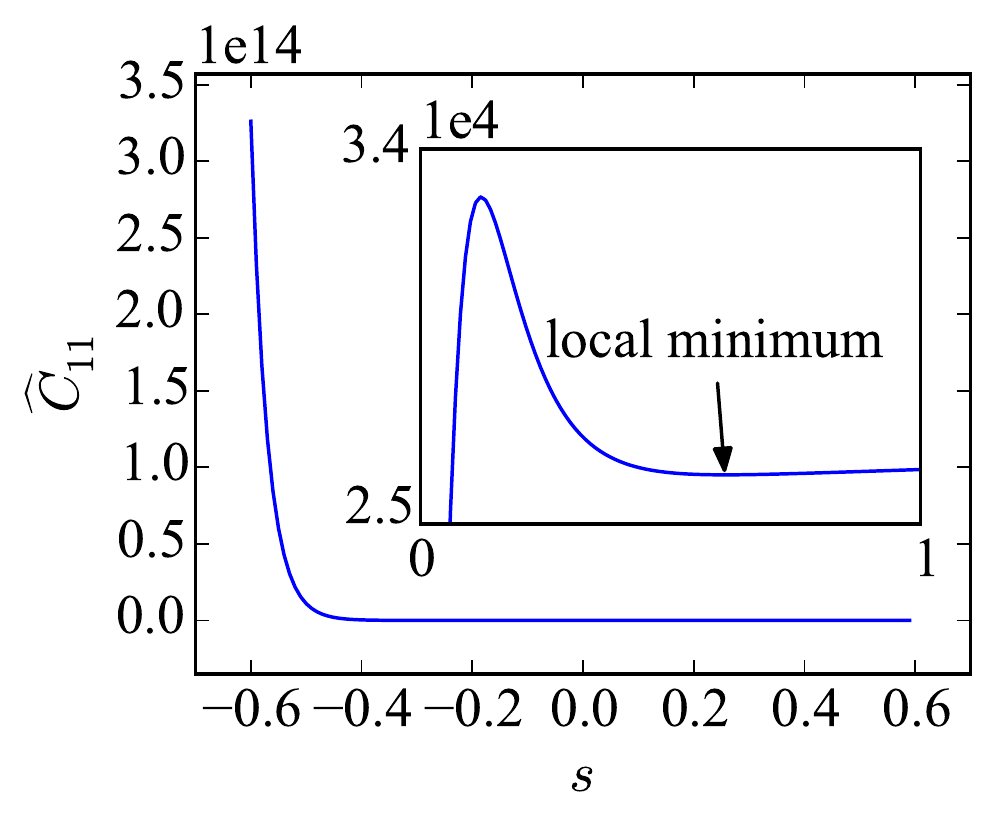}
\end{subfigure}
\caption{Cost along eigendirections. $\widehat{C}_1$ is associated with the smallest eigenvalue, and $\widehat{C}_{11}$ is associated with the largest. For $\widehat{C}_{10}$ and $\widehat{C}_{11}$ the insets show a zoomed view of the flat portions of the curves.}
\label{fig:cost_vs_alpha}
\end{figure}

The results in the previous section support the conjecture that the observed scatter in results is due to sloppy directions. However, since the Hessian is a local measure of the structure of $C$ near the optimal solution, it is possible that there are other minima further away that can trap the optimization process. To explore this possibility, we plot the cost function along the different eigendirections,
\begin{equation}
\widehat{C}_i(s) = C(\boldsymbol{\theta}^0 + s \boldsymbol{\Lambda}_i),
\end{equation}
where as before $\boldsymbol{\theta}^0$ corresponds to the original EDIP parameters and $s$ is a scalar factor that can be positive or negative.  The results are plotted in \fref{fig:cost_vs_alpha} for all eigendirections from the smallest eigenvalue 1 to the largest eigenvalue 11.  The cost $\widehat{C}$ varies more slowly along the eigendirections associated with the five smallest eigenvalues than along other directions, i.e.\ the cost function is more flat in these five directions.  (Note that the scale of the $y$-axis is not the same in the plots, and the first five plots have a significantly smaller range than the others.) This is in agreement with the previous observation that the spread of the fitted parameters is larger along these directions.  We also see that an additional local minimum (aside from the one associated with original parameters at $s=0$) is only apparent along the positive 11 direction at a significantly higher cost function.\footnote{We verified that this is a local minimum by checking the eigenvalues of the Hessian at the minimum and finding that they are all positive.} This minimum can trap the optimization algorithm for initial guesses in its vicinity, however over a large range near the original parameters there is only a single minimum. This suggests that a local minimizer is sufficient for this IP (using different initial guesses), which can reduce the fitting time drastically compared with global minimization efforts.

\subsection{Performance of minimization algorithms}
\label{sec:perform:min:methods}
Having established in the previous section that a local minimization is sufficient for EDIP (at least in the vicinity of the original parameters), we explore the efficiency of Powell's method, the geodesic LM algorithm added to \potfit, and the standard LM algorithm described in \cite{lourakis04LM} (see \sref{sec:glm} for details of LM and geodesic LM). The three minimizers are applied to fit EDIP. The initial guesses for the parameters are obtained in the same way as discussed in \sref{sec:sensitivity:param}. Three sets of 100 initial guesses are constructed by perturbing the original EDIP parameters by random numbers drawn from a Gaussian distribution with zero average and standard deviations of 0.1, 0.2 and 0.3. The results are given in \tref{tab:perform:min} showing for each minimizer and perturbation magnitude the number of attempts (out of 100) that converged below the cost level indicated on the left. For example for Powell's method with a perturbation magnitude of 0.2, the algorithm converged to a cost below $10^{-3}$ for 46 out of the 100 initial guesses. We see that geodesic LM outperforms the other two algorithms for all perturbation amplitudes. In all cases it is able to converge from more initial guess to below the specified tolerance. The standard LM algorithm is superior to Powell in all cases except for the smallest perturbation.

\begin{table}
\caption{\label{tab:perform:min} Statistics on the optimization of EDIP. The number of minimization attempts (out of a 100) that converge to a cost function below the tolerance specified on the left for the three minimizers studied and the three perturbation amplitudes (0.1, 0.2 and 0.3) used to generate the initial guesses.}
\begin{indented}
\item[]
\fl \begin{tabular}{@{\extracolsep{4pt}}l c c c c c c c c c}
\br
   & \multicolumn{3}{c}{Powell}   & \multicolumn{3}{c}{Standard LM}   & \multicolumn{3}{c}{Geodesic LM}\\
\cline{2-4}
\cline{5-7}
\cline{8-10}
Cost($C$)   & 0.1  & 0.2  & 0.3	     & 0.1  & 0.2  & 0.3      & 0.1  & 0.2  & 0.3\\
\mr
  $C < 10^{-7}$    & 85  & 37  & 7    & 68   & 46  & 9    & 100  & 60  & 14\\
  $C < 10^{-5}$    & 85  & 39  & 8	  & 76   & 49  & 9    & 100  & 60  & 14\\
  $C < 10^{-3}$    & 89  & 47  & 9	  & 100  & 58  & 16   & 100  & 64  & 21\\
  $C < 10^{-1}$    & 93  & 68  & 23	  & 100  & 79  & 25   & 100  & 84  & 33\\
  $C < 10^{0}$     & 98  & 78  & 36	  & 100  & 85  & 39   & 100  & 90  & 51\\
  $C < 10^{1}$     & 100 & 94  & 65	  & 100  & 96  & 54   & 100  & 98  & 76\\
\br
\end{tabular}
\end{indented}
\end{table}
\begin{figure}
\centering
\begin{subfigure}{0.49\columnwidth}
\includegraphics[width=\columnwidth]{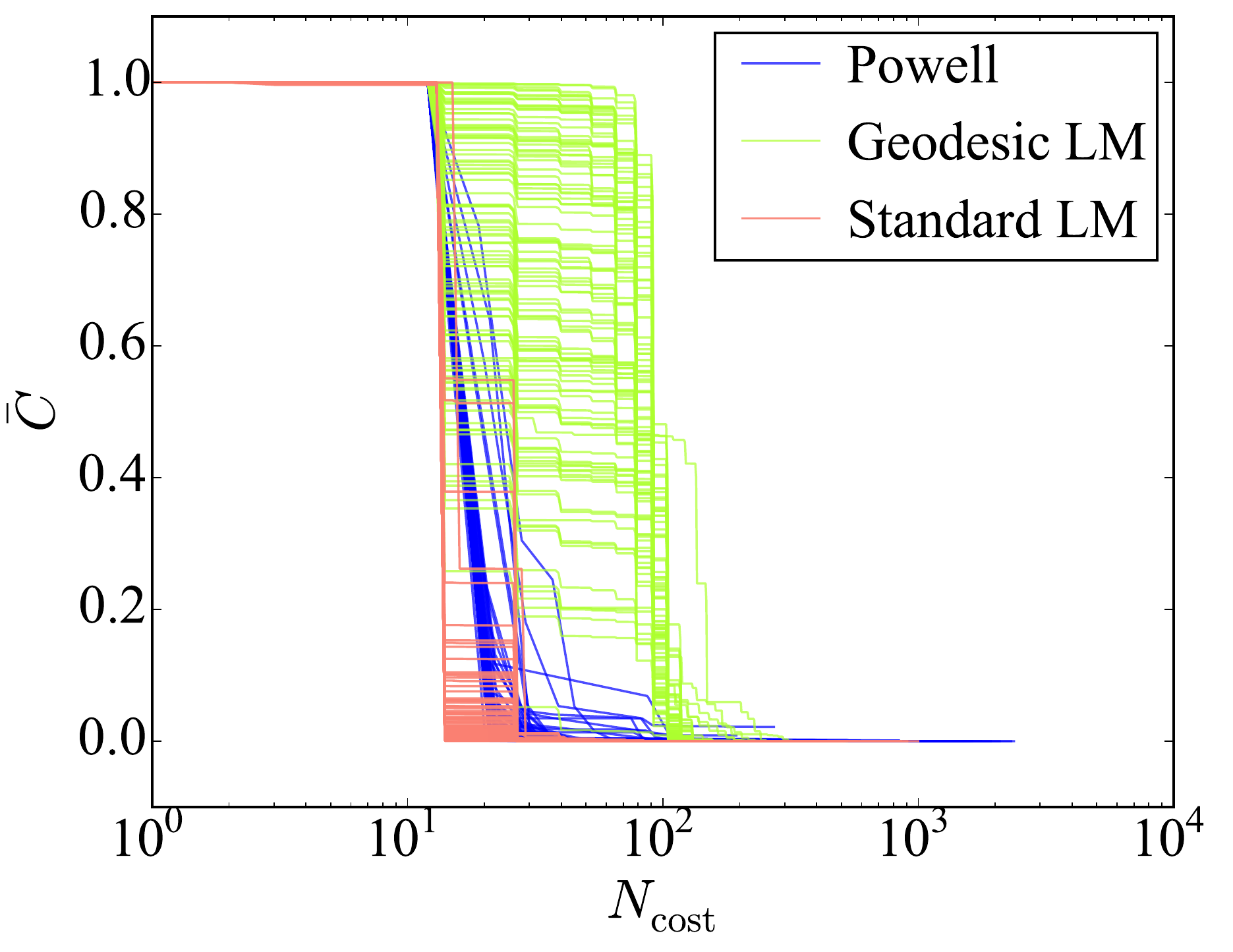}
\caption{}
\end{subfigure}
\begin{subfigure}{0.49\columnwidth}
\includegraphics[width=\columnwidth]{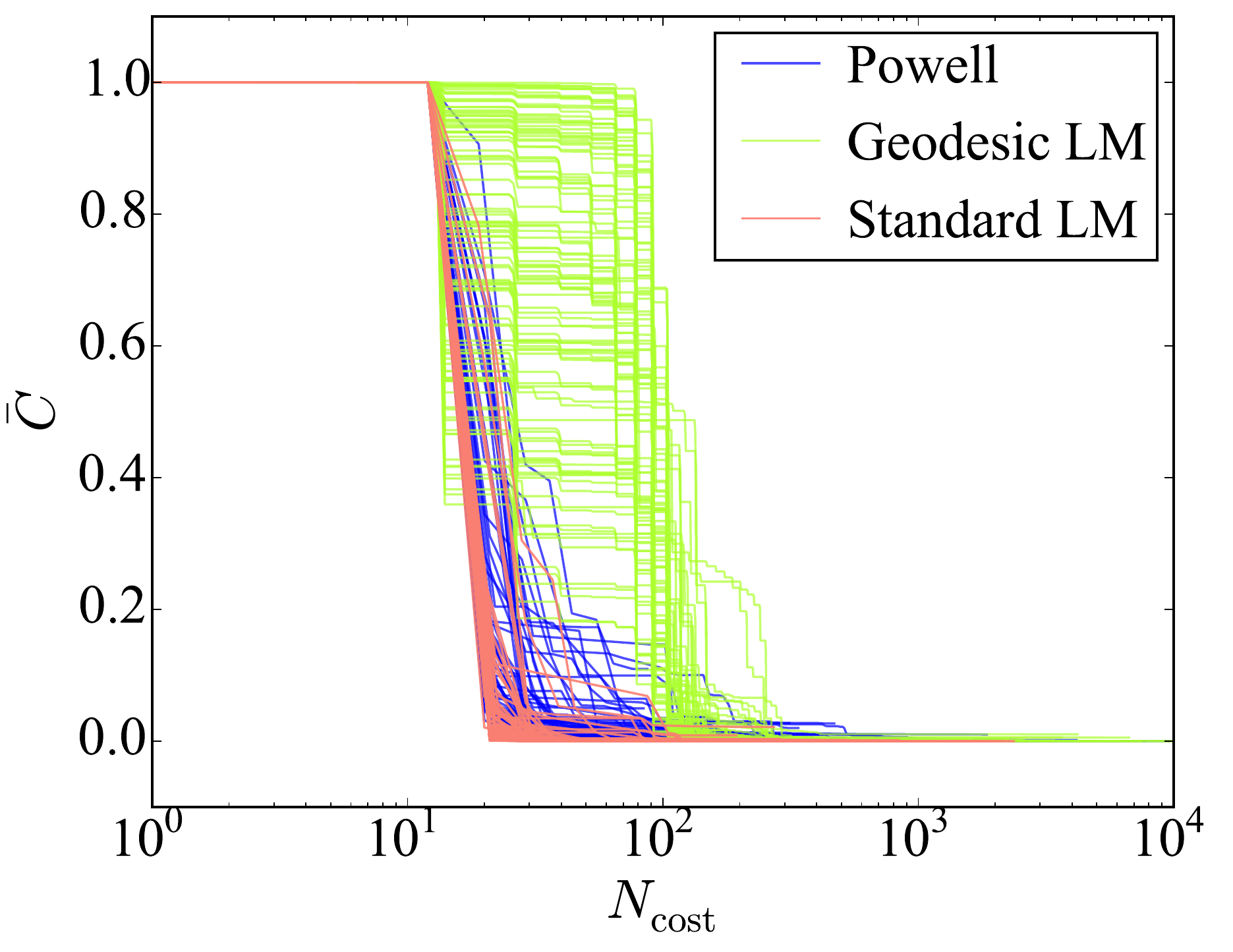}
\caption{}
\end{subfigure}
\begin{subfigure}{0.49\columnwidth}
\includegraphics[width=\columnwidth]{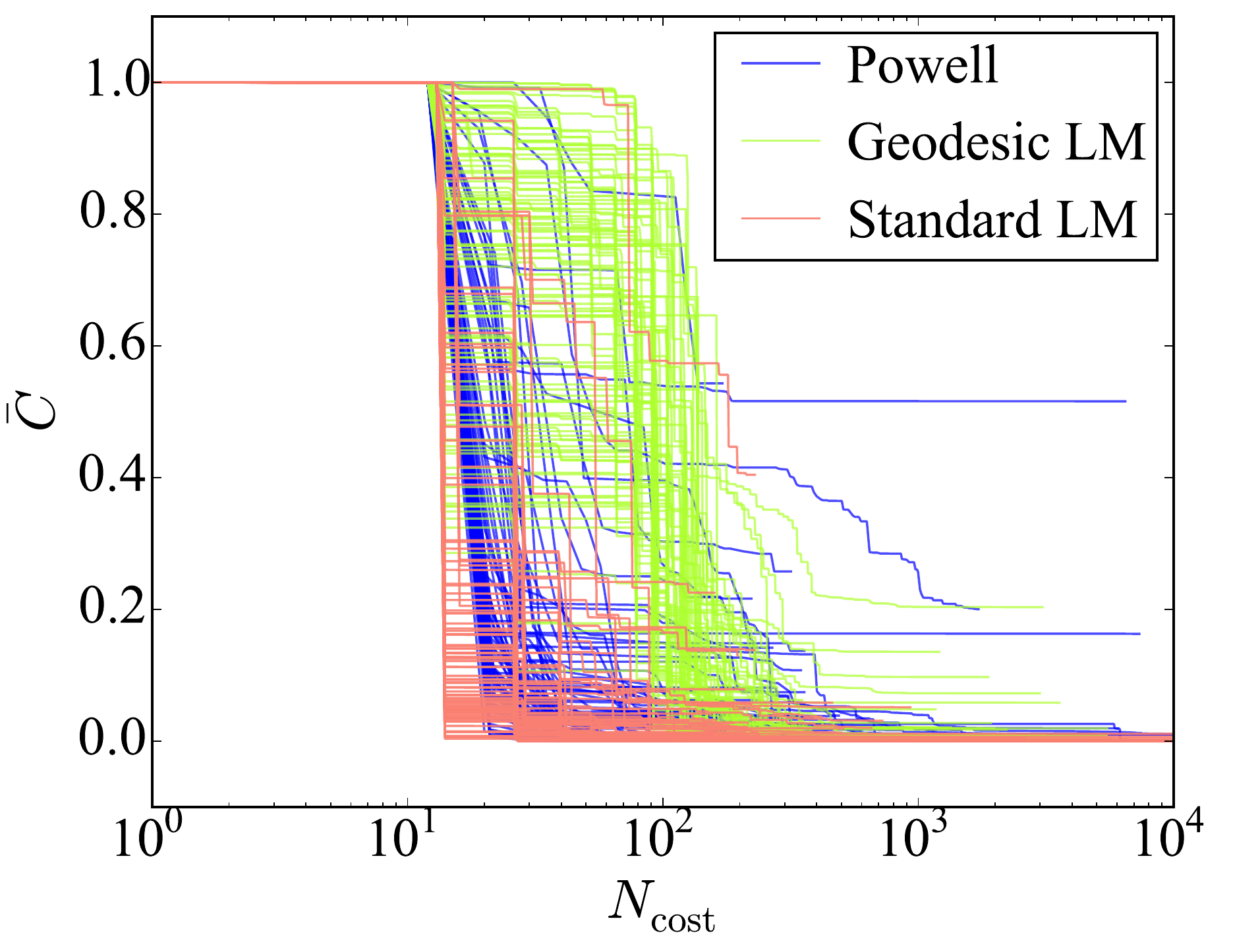}
\caption{}
\end{subfigure}
\caption{Normalized cost $\bar{C}$ versus the number of cost function evaluations $N_\text{cost}$. The initial guesses for the parameters are perturbations off the original EDIP parameters drawn from a Gaussian distribution with zero average and standard deviations of (a) 0.1, (b) 0.2, and (c) 0.3.}
\label{fig:cost:cevals}
\end{figure}

We are also interested in the computational cost of the different minimization methods. The Powell algorithm requires only evaluations of the cost function, whereas LM and geodesic LM require the gradient of the cost function with respect to the parameters. The gradient is computed using numerical differentiation, which only requires evaluations of the cost function. Therefore the number of cost function evaluations ($N_\text{cost}$) performed during the minimization is a good metric for comparing the performance of all three methods.  \Fref{fig:cost:cevals} shows for each perturbation amplitude, the normalized cost function $\bar{C} = C/C_\text{init}$ (where $C_\text{init}$ is the cost at the start of the minimization) versus $N_\text{cost}$ for the three methods. The initial flat region is associated with the setup stage of the methods (e.g. building an initial estimate for the Hessian in Powell's method).  Examining the plots, we see that Powell and LM are comparable in performance, whereas geodesic LM is about an order of magnitude slower. Thus geodesic LM's superior convergence properties come at the expense of a significant increase in the computational burden. For the current problem this tradeoff is acceptable since geodesic LM converges in most cases in less than 1000 cost function evaluations, which is quite reasonable for the current problem. However, when computation of the cost function is expensive (due to a larger training set or more complex IP) or when more steps are needed to converge as would be expected for IPs with more parameters, the computational cost of geodesic LM may be an issue.

\section{Summary and future directions}
\label{sec:summary}

The accuracy of an IP is critical to atomic-scale simulations.  Designing a high-quality transferable IP is a very challenging task due to the complexity of the quantum mechanical behavior that the IP is attempting to mimic.  The developer of an IP must select a functional form with adjustable parameters and a suitably-weighted training set of first principles and experimental reference data.  The IP is then optimized with respect to its parameters to reproduce the training set as closely as possible in terms of specified cost function. It is then validated against an independent test set of reference data.  If the validation results are inadequate, the functional form and/or the training set must be amended and the process repeated until a satisfactory IP is obtained.  This process is made more difficult by the fact that many IPs are sloppy models in the sense that their predictions are weakly dependent on certain combinations of their parameters.  This can cause optimization algorithms to fail to converge.

\potfit is an IP fitting program based on the force-matching method.  The idea is to include in the training set the forces on atoms computed from first principles for a set of atomic configurations of interest. This allows the developer to train the IP with a large amount of reference data spanning a larger portion of configuration space. We have extended \potfit in two ways:
\begin{enumerate}
\item We have incorporated a new optimization method based on the geodesic LM algorithm. Although slower than some other local minimizers it has improved convergence properties for sloppy models. This is the first application of geodesic LM to IP optimization.
\item We have adapted \potfit to be compatible with the KIM API. This allows \potfit to optimize KIM Models stored in the KIM Repository at \url{https://openkim.org}. This greatly expands the functional forms that \potfit can optimize.  The resulting IPs are portable and can be used seamlessly with many major software packages that support the KIM API.
\end{enumerate}

We study the effectiveness of the new code by using it to optimize the EDIP model for silicon. This IP is not available within the original \potfit code, but is available as a KIM Model.  We use EDIP to generate a training set of atomic configurations and test whether \potfit can reproduce the EDIP parameters from different initial guesses.  We make several observations:
\begin{enumerate}
\item EDIP displays the characteristics of a sloppy model.  For different initial guesses we find a large scatter in the parameter values.  An analysis shows that the scatter is maximal along the eigendirections of the Hessian of the cost function associated with the smallest eigenvalues. Thus the IP's behavior depends primarily on a few stiff combinations of parameters.
\item At least in the vicinity of the original parameters the cost function is largely convex with only two local minima. This suggests that a local minimization method with different random initial guesses is sufficient.
\item The geodesic LM algorithm was on average twice as likely to converge to the correct solution from different initial guesses than standard LM or Powell's method that tended to get trapped along sloppy directions.  However, the improved performance comes at the cost of about an order of magnitude increase in computational expense for geodesic LM.  For the current problem the computations were not prohibitive and geodesic LM was the preferred method.
\end{enumerate}

The extended \potfit code is the first step in developing an IP fitting framework based on the OpenKIM project.  Such a framework would make it possible for the developer to draw upon the large selection of KIM Tests (standardized codes for computing material properties) and reference data in OpenKIM. Models developed using this framework \emph{and} the training set used in the fit will then be archived at \url{https://openkim.org}. The latter is an important advantage as it means that other developers will be able to draw upon previous work and adapt a Model to a new application starting from the training set used in its development. This framework is currently under development for use in fitting IPs for 2D materials \cite{NovoselovPNAS05}.

\ack
We thank Mark Transtrum for his open source implementation of the geodesic LM algorithm, and Colin Clement for useful discussions.  The authors also thank Daniel Schopf for his assistance in merging the OpenKIM extension into the main \emph{potfit} distribution. The research was partly supported through the National Science Foundation (NSF) under grants No.\ PHY-0941493 and DMR-1408211. The authors acknowledge the support of the Army Research Office (W911NF-14-1-0247) under the MURI program.

\section*{References}

\bibliographystyle{iop}
\bibliography{kim-potfit}

\end{document}


\title[Supplementary information for ``A {KIM}-compliant \potfit for fitting sloppy IPs'']{Supplementary information for ``A {KIM}-compliant \potfit for fitting sloppy interatomic potentials: {A}pplication to the {EDIP} model for silicon''}

\author{Mingjian~Wen, Junhao~Li, Peter~Brommer, Ryan~S.~Elliott, James~P.~Sethna and Ellad~B.~Tadmor\footnote{Corresponding author.}}
\ead{tadmor@umn.edu}

\address{Department of Aerospace Engineering and Mechanics, University of Minnesota, Minneapolis, MN 55455, USA}
\address{$^{1}$ Warwick Centre for Predictive Modelling, School of
 Engineering, and Centre for Scientific Computing, University of Warwick, Coventry CV4~7AL, UK}

\maketitle

\section{The EDIP potential for silicon}

In this section we describe the functional form of the Environment Dependent Interatomic Potential (EDIP) for bulk silicon developed by Bazant et al.\  \cite{bazant1996modeling, bazant1997environment, justo1998edip}.  The total potential energy of a configuration of $n$ atoms is assumed to take the form $E = \sum_{i=1}^n E_i$, where the contribution to the energy of atom $i$ is given by
\begin{equation}
 E_i = \sum_{j \neq i} V_2(R_{ij},Z_i) + \sum_{j \neq i} \sum_{k \neq i, k > j} V_3 (\vec{R}_{ij},\vec{R}_{ik},Z_i) .
\end{equation}
The ``local environmental dependence'' of atom $i$ is introduced through its effective coordination number, defined by,
\begin{equation}
Z_i = \sum_{m \neq i} f(R_{im}) ,
\end{equation}
where $f(R_{im})$ is a cutoff function given by,
\begin{equation}
f(r) = \cases{1 & if $r < c$\\
       \exp{\left(\frac{\alpha}{1-x^{-3}}\right)} & if  $c < r < a$\\
       0 & if  $r > a$\\}
\end{equation}
in which $ x = ( r-c )/(a-c) $.

The two-body term $V_2$ includes a repulsive part and an attractive part:
\begin{equation}
V_2(r,Z) = A \left[ \left(\frac{B}{r}\right)^{\rho} - p(Z) \right] \exp\left( \frac{\sigma}{r-a} \right)
\end{equation}
with the attractive bond-order function taking the form $p(Z) = e^{-\beta Z^2}$.

The three-body term $V_3$ takes the form of a product of radial and angular functions:
\begin{equation}
V_3(\vec{R}_{ij}, \vec{R}_{ik}, Z_{i}) = g(R_{ij})g(R_{ik}) h(l_{ijk},Z_i) .
\end{equation}
The radial contribution to $V_3$ is given by $g(r) = \exp(\gamma/(r-a))$, and the angular contribution is
\begin{equation}
 h(l,Z) = \lambda \left[ \left( 1 - e^{-Q(Z)(l+\tau(Z))^2}\right)  + \eta Q(Z)\left(l+\tau(Z)\right)^2\right]
\end{equation}
with the bond angle dependence contained in $l_{ijk} = \cos \theta_{ijk} = \vec{R}_{ij} \cdot \vec{R}_{ik} / R_{ij} R_{ik}$ and
the coordination number dependence controlled by the function $Q(Z) = Q_0 e^{-\mu Z}$.  The function $\tau(Z)$, which controls the equilibrium angle of the three-body interaction, is chosen to have the form
\begin{equation}
 \tau(Z) = u_1 + u_2 \left( u_3 e^{-u_4 Z} - e^{-2 u_4 Z} \right),
\end{equation}
where $u_1=-0.165799$, $u_2=32.557$, $u_3=0.286198$ and $u_4=0.66$ are parameters chosen such that $\tau(Z)$ yields correct theoretical bond angle at $Z = 2,3,4$, and $6$, and interpolates smoothly between them.

The EDIP potential has 13 adjustable parameters: $A$, $B$, $\rho$, $\beta$, $\sigma$, $\lambda$, $\eta$, $\gamma$, $\mu$, $\alpha$, $Q_0$, $a$, and $c$, and their values are listed in \tref{tab:edip:param:val}. The thresholds values $a$ and $c$ used in the cutoff function $f(r)$ are fixed during the fitting process.  Consequently, only 11 parameters are optimized.

\begin{table}
\caption{EDIP Parameter values for silicon}
\label{tab:edip:param:val}
\vspace*{10pt}
\centering
\begin{tabular}{ccc}
\hline
\hline
Parameter   &Value          \\
\hline
$A$         &7.9821730 eV   \\
$B$         &1.5075463 \AA  \\
$\rho$      &1.2085196      \\
$\beta$     &0.0070975      \\
$\sigma$    &0.5774108 \AA  \\
$\lambda$   &1.4533108 eV   \\
$\eta$      &0.2523244      \\
$\gamma$    &1.1247945 \AA  \\
$\mu$       &0.6966326      \\
$\alpha$    &3.1083847      \\
$Q_0$       &312.1341346    \\
$a$         &3.1213820 \AA  \\
$c$         &2.5609104 \AA  \\
\hline
\hline
\end{tabular}
\end{table}

\section{Training set}

The training set consists of the energy and forces computed with EDIP for a periodic configuration of 1000 atoms perturbed off the ideal diamond structure positions.  The training set file ``supplementary\_traning\_set.txt'' accompanies this article. The header of the file specifies the supercell lattice vectors and the average energy per atom. This is followed by a list giving for each atom the atomic species code, atom coordinates, and force on the atom. See the \potfit webpage \url{http://www.potfit.net/wiki/doku.php} for details on the format of the training set file.

\section*{References}

\bibliography{supplementary.bib}
\bibliographystyle{unsrt}